\documentclass[twocolumn,twocolappendix,usenatbib,useAMS,fleqn,usenames,dvipsnames, tighten]{aastex63}
\usepackage{amsmath}
\hypersetup{linkcolor=MidnightBlue,citecolor=MidnightBlue,filecolor=MidnightBlue,urlcolor=MidnightBlue}
\graphicspath{{./}{plots/}}
\accepted{ApJ}
\usepackage{enumitem,amsmath,amstext,mathtools}
\newcommand{\yt}{{\sc yt} }
\newcommand{\trident}{{\sc trident} }
\newcommand{\po}{{\sc P0}}
\newcommand{\pocr}{{\sc P0+CR}}

\newcommand{\civ}{C~{\sc iv}}
\newcommand{\si}{Si~{\sc iii}}
\newcommand{\hi}{H~{\sc i}}
\newcommand{\mgii}{Mg~{\sc ii}}
\newcommand{\ovi}{O~{\sc vi}}

\hypersetup{linkcolor=MidnightBlue,citecolor=MidnightBlue,filecolor=MidnightBlue,urlcolor=MidnightBlue}

\begin{document}

\shorttitle{CGM Kinematics with cosmic rays}
\shortauthors{Butsky, Werk, et al.}

\title{The Impact of Cosmic Rays on the Kinematics of the Circumgalactic Medium}
\correspondingauthor{Iryna S. Butsky}
\email{ibutsky@caltech.edu}

\author[0000-0003-1257-5007]{Iryna S. Butsky}
\affiliation{TAPIR, Mailcode 350-17, California Institute of Technology, Pasadena, CA 91125, USA}
\affiliation{Astronomy Department, University of Washington, Seattle, WA 98195, USA}
\author[0000-0002-0355-0134]{Jessica K. Werk}
\affiliation{Astronomy Department, University of Washington, Seattle, WA 98195, USA}
\author{Kirill Tchernyshyov}
\affiliation{Astronomy Department, University of Washington, Seattle, WA 98195, USA}
\author[0000-0003-3806-8548]{Drummond B. Fielding}
\affiliation{Center for Computational Astrophysics, Flatiron Institute, 162 Fifth Avenue, New York, NY 10010, USA}
\author{Joseph Breneman}
\affiliation{Rutgers University, Department of Physics and Astronomy, 136 Frelinghuysen Road, Piscataway, NJ 08854, USA}
\affiliation{Astronomy Department, University of Washington, Seattle, WA 98195, USA}
\author{Daniel R. Piacitelli}
\affiliation{Astronomy Department, University of Washington, Seattle, WA 98195, USA}
\author{Thomas R. Quinn}
\affiliation{Astronomy Department, University of Washington, Seattle, WA 98195, USA}
\author{N. Nicole Sanchez}
\affiliation{Astronomy Department, University of Washington, Seattle, WA 98195, USA}
\author{Akaxia Cruz}
\affiliation{Physics Department, University of Washington, Seattle, WA 98195, USA}
\author[0000-0002-3817-8133]{Cameron B. Hummels}
\affiliation{TAPIR, California Institute of Technology, Pasadena, CA 91125, USA}
\author[0000-0002-1979-2197]{Joseph N. Burchett}
\affiliation{Department of Astronomy, New Mexico State University, Las Cruces, NM 88003, USA}
\author[0000-0002-4353-0306]{Michael Tremmel}
\affiliation{Department of Physics, Yale University, New Haven, CT 06511, USA}
\affiliation{Yale Center for Astronomy and Astrophysics, New Haven, CT 06511, USA}


\begin{abstract}
We use hydrodynamical simulations of two Milky Way-mass galaxies to demonstrate the impact of cosmic-ray pressure on the kinematics of cool and warm circumgalactic gas. Consistent with previous studies, we find that cosmic-ray pressure can dominate over thermal pressure in the inner 50 kpc of the circumgalactic medium (CGM), creating an overall cooler CGM than that of similar galaxy simulations run without cosmic rays. We generate synthetic sightlines of the simulated galaxies' CGM and use Voigt profile fitting methods to extract ion column densities, Doppler-$b$ parameters, and velocity centroids of individual absorbers. We directly compare these synthetic spectral line fits with HST/COS CGM absorption-line data analyses,  which tend to show that metallic species with a wide range of ionization potential energies are often kinematically aligned. Compared to the Milky-Way simulation run without cosmic rays,  the presence of cosmic-ray pressure in the inner CGM  creates narrower \ovi\ absorption features and broader \si~ absorption features, a quality which is more consistent with observational data. Additionally, because the cool gas is buoyant due to nonthermal cosmic-ray pressure support, the velocity centroids of both cool and warm gas tend to align in the simulated Milky Way with feedback from cosmic rays. Our study demonstrates that detailed, direct comparisons between simulations and observations, focused on gas kinematics, have the potential to reveal the dominant physical mechanisms that shape the CGM.

\end{abstract}

\section{Introduction}

Galaxies evolve embedded within a vast, gaseous halo that dwarfs the mass and spatial extent of the galactic disk. This circumgalactic medium (CGM) has a rich multiphase structure with gas temperatures, densities, and metallicities spanning several orders of magnitude \citep{Tumlinson:2017}. The CGM drives galaxy evolution by controlling gas accretion, which is necessary for continued star formation, and is in turn shaped by galactic outflows, which expel the byproducts of stellar evolution. Constraints on the structure of the CGM and the origin of its multiple phases are therefore crucial for understanding the dominant physical mechanisms driving galaxy evolution. 

Because the CGM is so diffuse, much of what we know about its composition and kinematics derives from absorption-line studies, in which the sightline to a bright background source (typically a quasar) intersects a galaxy's CGM. The atoms in the CGM interact with the light from the quasar, creating absorption-line features in the resulting spectra from which we can measure properties like the ionic column density and line-of-sight velocity of the absorbing gas. Most sightlines intersecting the CGM  detect absorption from high-, intermediate-, and low-ionization species at rest-frame ultraviolet (UV) wavelengths, spanning over an order of magnitude in ionization potential energies, both at low redshift \citep[e.g., z $\lesssim$ 1: ][]{Bergeron:1991qy, Prochaska:2011, Tripp:2011, Tumlinson:2013, Werk:2013, Nielsen:2013, Bordoloi:2014, Liang:2014, Borthakur:2015, Burchett:2019} and high redshift \citep[e.g.,  $2 < z < 3$: ][]{Steidel:2010, Rudie:2012, Turner:2015, Zahedy:2019}. Absorption from low-ions (e.g., \mgii, \si), assumed to trace cool $10^{4-5}$ K gas, is prevalent in the inner $\sim 100$ kpc of the CGM while absorption from intermediate and high ions (e.g., \civ, \ovi), assumed to trace warm $10^{5-6}$ K gas, is detected out to, and sometimes beyond the galaxy virial radius \citep{Johnson:2015, Burchett:2016, Keeney:2018}. Kinematic studies reveal that high ions and low ions often have similar line-of-sight velocity centroids, suggesting a common (or related) physical distribution of the warm and cool absorbers \citep{Tripp:2008, Tripp:2011, Burchett:2015, Werk:2016, Haislmaier:2021}. 

Ionization modeling of the measured quantities has yielded a rich set of constraints on the gas-phase metallicity of the CGM \citep[e.g.,][]{Lehner:2013, Prochaska:2017, Lehner:2019}, its total baryonic content \citep[e.g.,][]{Werk:2014, Stern:2016}, and its pressure profile \citep[][]{Stocke:2013, Werk:2016, Voit:2019a}. In particular, around low-redshift L$^*$ galaxies, cool CGM gas appears to have significantly lower densities than required for it to be in thermal pressure equilibrium with the hot phase \citep{Werk:2014, Stern:2016}. 

One promising interpretation of the low densities of cool gas is that the CGM is supported by nonthermal cosmic-ray pressure. In recent years, cosmic rays have been invoked in galaxy simulations to launch winds \citep{Ipavich:1975, Uhlig:2012, Booth:2013, Pakmor:2016, Simpson:2016, Ruszkowski:2017, Farber:2018, Hopkins:2021_outflows} and thereby alter the ionization structure of the CGM \citep{Salem:2016, Butsky:2018, Ji:2020, Buck:2020}. Although the quantitative details of the predicted CGM structure depend on the invoked model of cosmic-ray transport \citep{Butsky:2018, Hopkins:2021_transport1, Hopkins:2021_transport2}, which is still poorly constrained, many of these simulations predict a cosmic-ray-pressure-supported CGM around low-redshift L* galaxies. In these models, the cosmic-ray pressure can exceed the gas pressure by up to two orders of magnitude in the inner CGM. Significant cosmic-ray pressure support allows low-density cool gas to survive in the CGM, naturally producing high column densities of low ions. Significant cosmic-ray pressure can also qualitatively alter the kinematics of the CGM by changing the morphology of outflows \citep[e.g.,][]{Girichidis:2016, Jana:2020_outflow} and preventing a virial shock \cite{Ji:2021}.

In this work, we explore the implications of a cosmic-ray-pressure-supported halo on the observed kinematic alignment between multiphase ions. We start with a Milky Way mass galaxy which has successfully reproduced \ovi\ column densities with the help of supermassive black hole (SMBH) feedback \citep{Sanchez:2019}. We then re-simulate that same galaxy, adding supernova cosmic-ray feedback. By redshift $z = 0.25$, this galaxy develops a cosmic-ray-pressure-supported halo, qualitatively similar to those described in the recent literature. 

To compare the simulations with observations, we generate 100 synthetic spectra from sightlines that pierce the inner CGM of each simulated halo. We then analyze absorption features in these spectra with Voigt-profile fitting tools used by observers. This approach allows us to directly compare our simulated CGM against observed absorption-line properties (e.g., Doppler-$b$ parameters, line-of-sight velocity offsets) --- an approach that is inaccessible to traditional simulation analysis techniques.

This paper is organized as follows. In \autoref{sec:methods}, we describe our methods, including the simulation initial conditions, our choices of run-time physics, and the process for generating and analyzing synthetic spectra. In \autoref{sec:results}, we present our results. We first present a simulation-oriented view of the CGM properties and then shift the focus to properties derived from the Voigt-profile fits with an emphasis on the Doppler-$b$ parameters and velocity offsets. In \autoref{sec:discussion}, we discuss the implications of our work and future directions. In \autoref{sec:summary}, we present a summary of this work and its conclusions. Lastly, we demonstrate the implementation of cosmic-ray physics in the {\sc ChaNGa} simulation code and perform additional tests in \autoref{sec:appendix_changa}, \autoref{sec:appendix_crtests}, and \autoref{sec:appendix_GM}.

\section{Methods}\label{sec:methods}
\subsection{Description of Galaxy Simulations}
In this work, we analyze cosmological zoom-in simulations of two Milky Way-sized ($M_{\rm vir} = 9.9\times 10^{11} M_{\odot}$ at $z = 0$) galaxies evolved with and without cosmic-ray feedback. The base model, ``Patient 0" (\po) has been studied in great detail in \citet{Sanchez:2019} and \citet{Sanchez:2020}. ``Patient 0 with cosmic rays" (\pocr) uses the same initial conditions and physics as \po\ except that it also includes cosmic-ray feedback from supernovae. We present a basic description of the simulation here for context, but encourage the reader to see the referenced works for more detail. The description of the cosmic-ray physics implementation is detailed in \autoref{sec:appendix_changa}.

The simulations were run with ChaNGa, a smoothed-particle hydrodynamics (SPH) astrophysical simulation library \citep{Menon:2015} with a recently updated formalism that includes a geometric density approach in the force expression \citep{Wadsley:2017}. We assume a $\Lambda$CDM cosmology with parameters $\Omega_m = 0.3086, \Omega_{\Lambda} = 0.6914, h = 0.67, \sigma_s = 0.77$ and an extragalactic UV background described by \cite{HaardtMadau:2012}. The mass of dark matter particles and gas particles are $2.1\times10^5 {\rm M}_{\odot}$ and  $1.4\times10^5 {\rm M}_{\odot}$ respectively.

We do not include metal-line cooling for temperatures above $10^4$ K. Although this is a limitation of the simulation, these effects are mitigated by the fact that the balance between heating and cooling at warm CGM temperatures is dominated by the extragalactic UV background. In this case, excluding metal-line cooling overestimates the cooling time by a factor of $\sim 3-5$ \citep{Shen:2010}. Additionally, the cooling times in $10^{4-4.5}$ K gas are well estimated by primordial cooling \citep{Hopkins:2018}. 

Both simulations model star formation assuming a Kroupa IMF \citep{Kroupa:2001} and `blastwave' supernova feedback \citep{Stinson:2006}. We allow star formation to happen in cold ($T < 10^4$ K), dense ($n > 0.2\, \mathrm{cm}^{-3}$) gas with an efficiency of 15\% and a characteristic timescale of $10^6$ years. Each supernova injects $0.75 \times 10^{51}$ erg of energy into the surrounding gas. Supernovae also inject $1.4 M_{\odot}$ of mass into all neighboring particles within the smoothing length. Of the total injected mass, $0.63 M_{\odot}$ is assumed to be iron and $0.13 M_{\odot}$ is assumed to be oxygen, based on SNIa yield models \citep{Thielemann:1986}. Metal transport is modeled by diffusion as described in \citet{Shen:2010}.

In \pocr, cosmic-ray energy is injected into the simulation during supernova feedback events. The total injected supernova energy remains constant, but 10\% is injected in the form of cosmic-ray energy. Cosmic-ray transport is modeled as isotropic diffusion with a constant diffusion coefficient of $\kappa = 10^{29} {\rm cm}^2 {\rm s}^{-2}$. 

We also include SMBH feedback as described in \citet{Tremmel:2017}. SMBH seeds form in very dense gas ($n > 3 m_H/{\rm cm}^3$) with low metallicities ($Z < 3 \times 10^{-4} Z_{\odot}$). SMBH accrete gas through both mergers and modified Bondi-Hoyle accretion \citep{Tremmel:2017} and redistribute 0.2\% of the mass energy of their accreted material into nearby gas via thermal feedback. We do not inject additional cosmic-ray energy during SMBH feedback events. 

\begin{figure}
\includegraphics[width=0.5\textwidth]{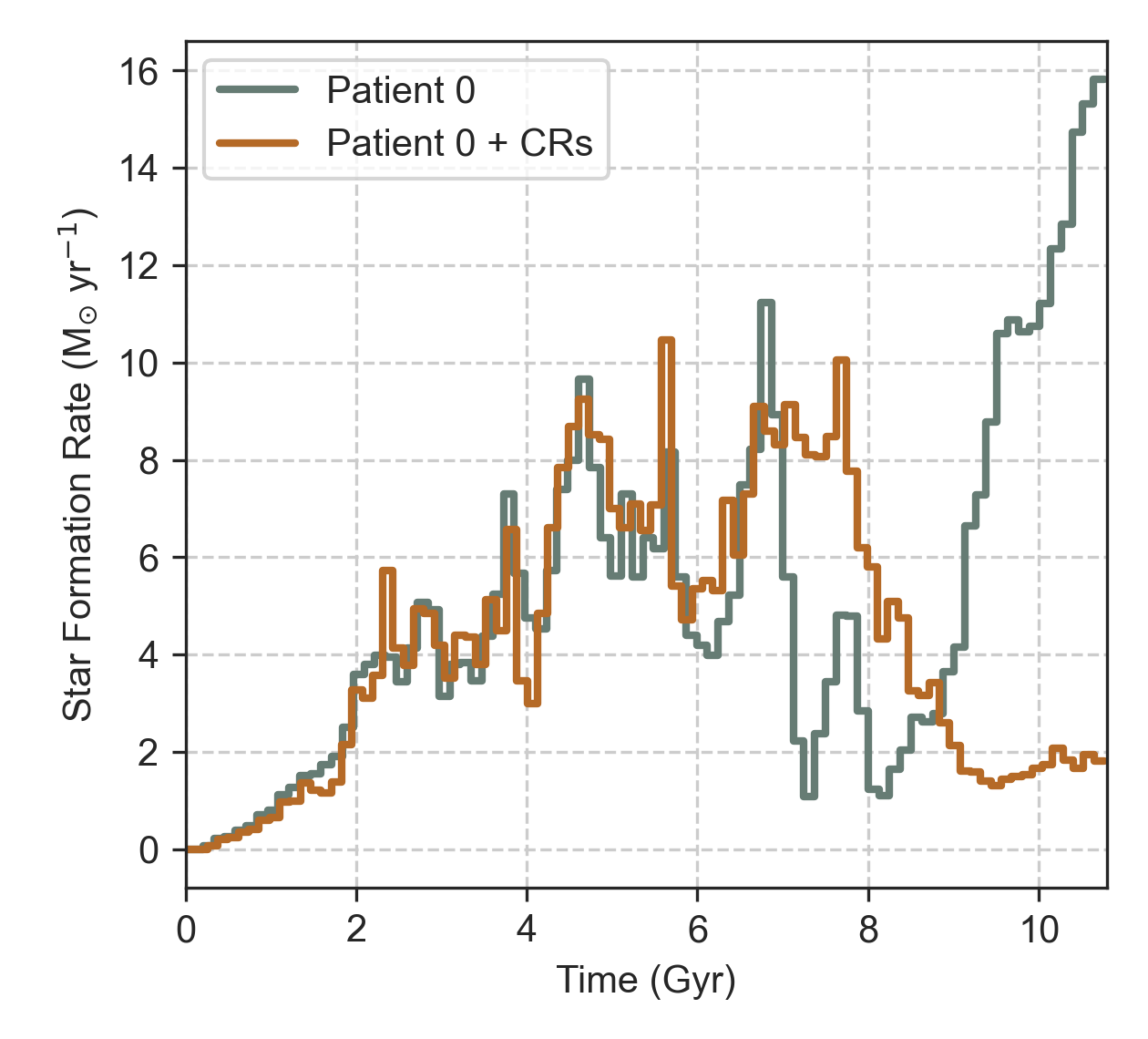}
\caption{The star formation rate over time for \po\ and \pocr. The two galaxies have nearly identical star formation histories for the first 6 Gyr of their evolution. As cosmic-ray pressure builds up in the galactic disk it, suppresses star formation in \pocr.} 
\label{fig:sfh} 
\end{figure}

We show the star formation histories of \po\ and \pocr\ in \autoref{fig:sfh}. Although the two galaxies share similar star formation histories early in their evolution, after $\sim 8$ Gyr, \pocr\ has a significantly reduced star formation rate relative to \pocr, caused by the build up of cosmic-ray pressure within the galactic disk. At redshift $z = 0.25$ (at which we perform our CGM analysis) the stellar masses of \po\ and \pocr\ are $5.5 \times 10^{10} M_{\odot}$ and $4.6\times 10^{10} M_{\odot}$ respectively.

\begin{figure*}
\includegraphics[width=0.5\textwidth]{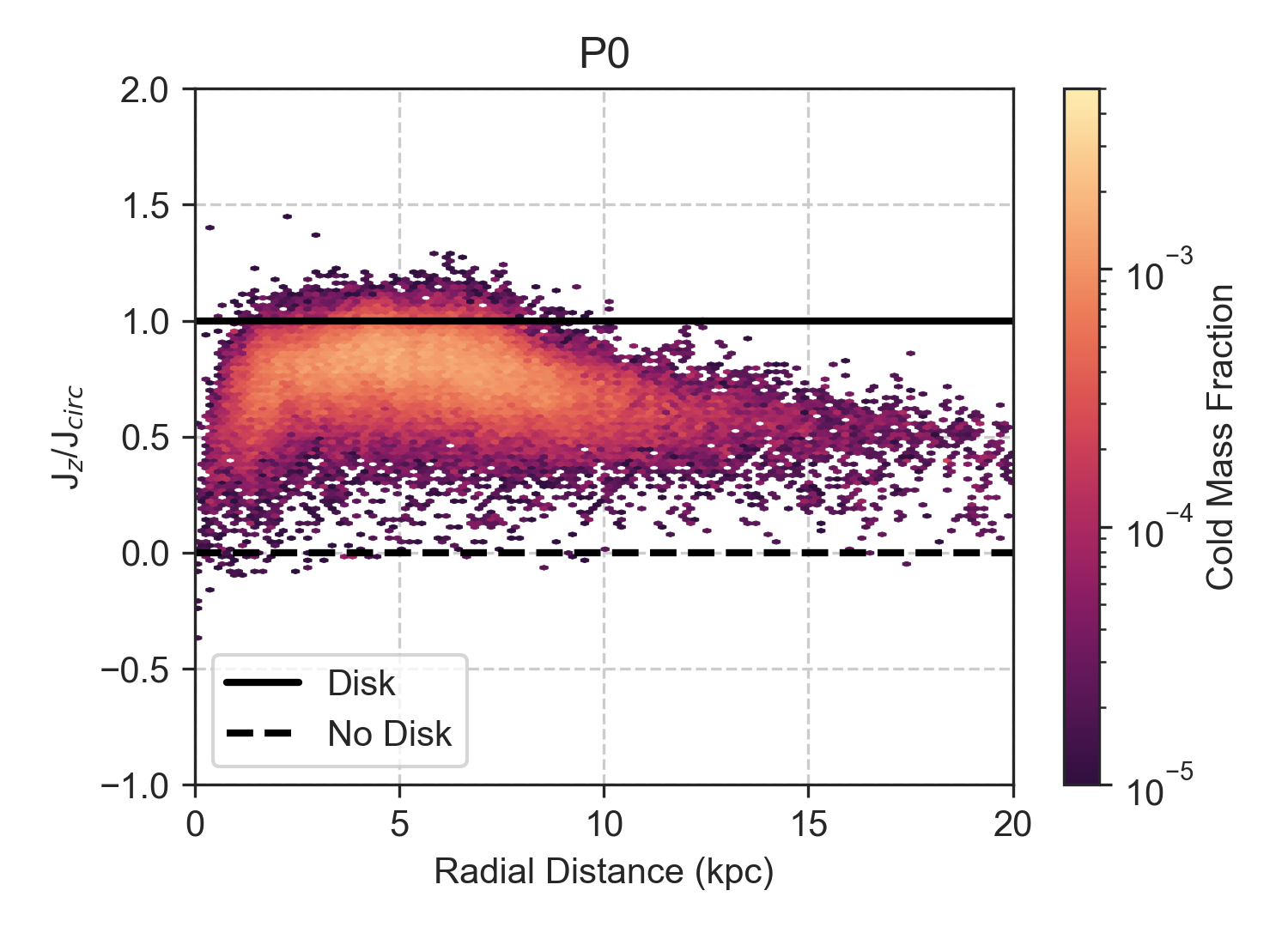}
\includegraphics[width=0.5\textwidth]{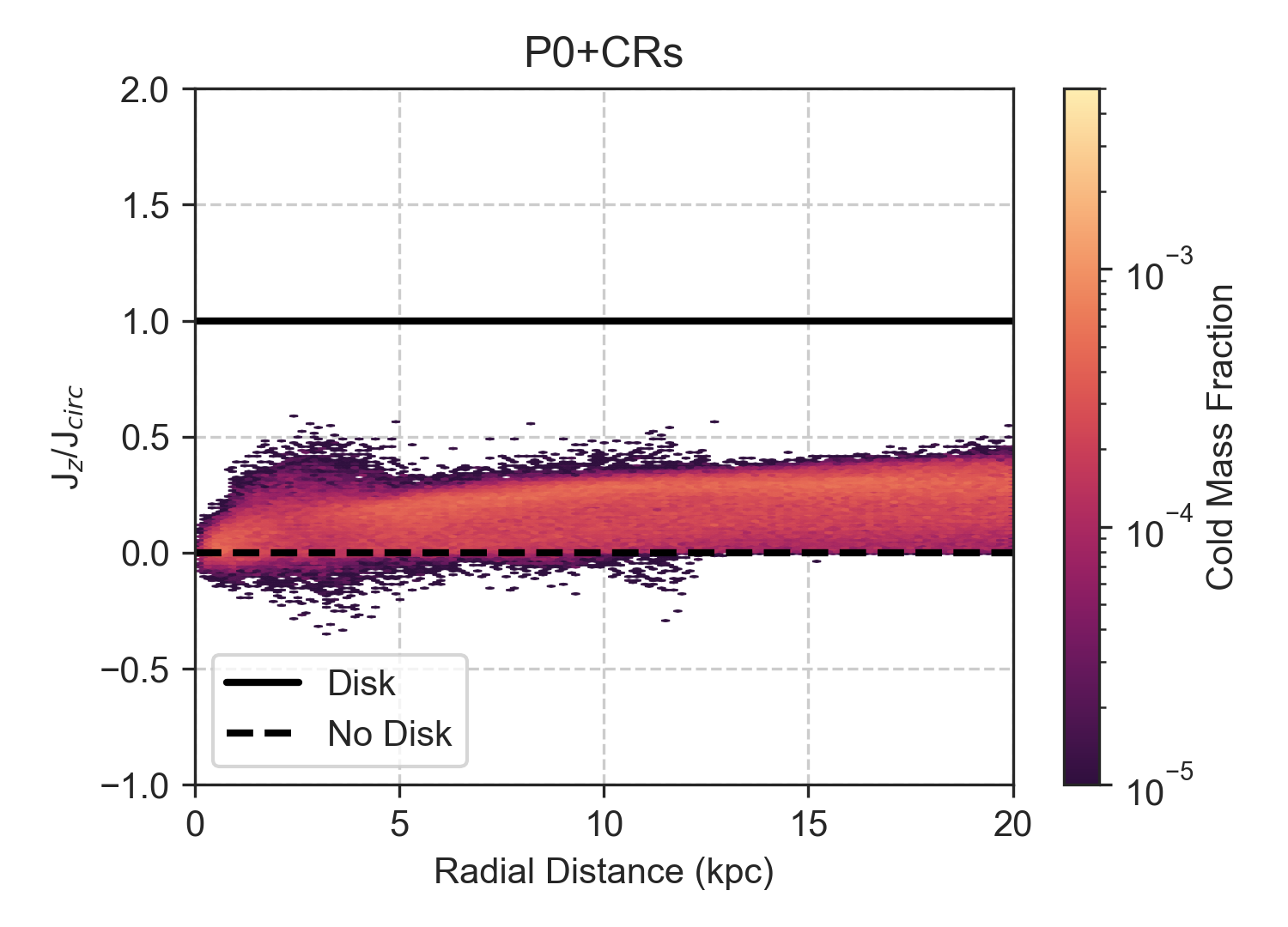}

\caption{The circularity parameter, $j_z / j_{\rm circ}$,  as a function of radial distance from the galaxy center, colored by the cold gas ($T < 2\times 10^4$ K) mass fraction. For each gas particle, the circularity parameter measures the specific angular momentum perpendicular to the disk, $j_z$, relative to a theoretical, perfectly circular specific angular momentum. Therefore, circularity parameters with $j_z / j_{\rm circ} \sim 1$ indicate the presence of rotationally-supported galactic disk (\po) while $j_z / j_{\rm circ} < 0.5$ is characteristic of a dispersion-dominated galaxy (\pocr). } 
\label{fig:circ} 
\end{figure*}

We highlight the differences in galaxy disk morphology with the circularity parameter, $j_z/j_{\rm circ}$ \citep{Keller:2015, Sanchez:2020}. For each gas particle within 20 kpc of the galaxy center, we calculate the specific angular momentum component perpendicular to the disk, $j_z$, and compare it against a theoretical perfectly circular specific angular momentum in that gravitational potential well, $j_{\rm circ}$. \autoref{fig:circ} demonstrates that while \po\ has a rotationally-supported disk ($j_z/j_{\rm circ} \approx 1)$, the ISM in \pocr\ is dispersion-dominated ($j_z/j_{\rm circ} < 0.5$). This difference in galactic disk morphology is enhanced in our simulations, likely due to the lack of Coulomb and hadronic losses, which results in an over-concentration of cosmic-ray pressure in the galactic disk. In general, galactic disk structure is sensitive to the details of cosmic-ray transport physics and simulations with substantial cosmic-ray pressure support in their CGM can range from having rotationally-supported to dispersion-dominated disks \citep[e.g.,][]{Chan:2019, Buck:2020, Ji:2020}.

\begin{figure}
\includegraphics[width=0.5\textwidth]{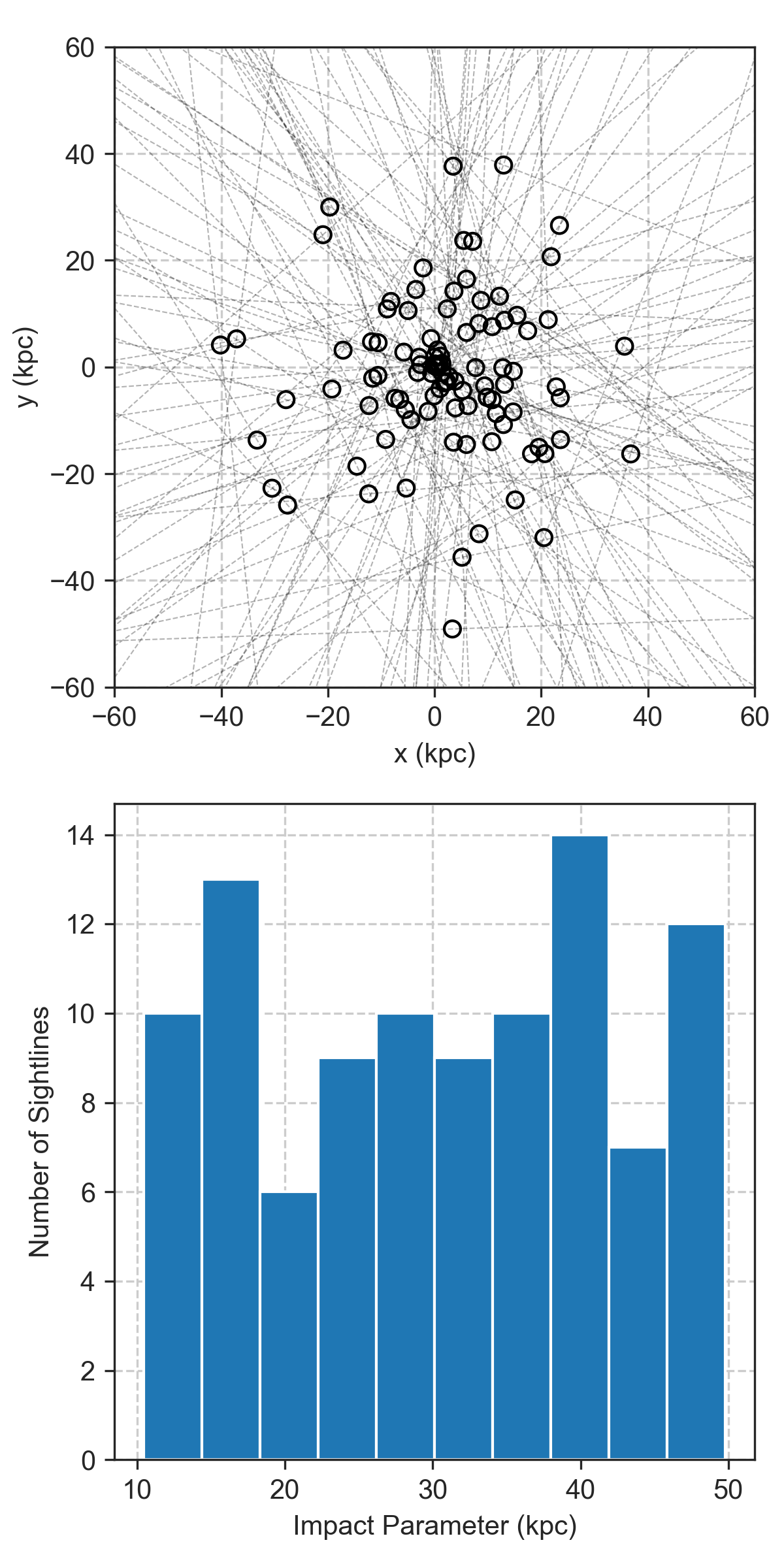}
\caption{ (Top) The projected paths of the 100 synthetic spectra in the x-y plane, centered on the galaxy center. The open circles show the projected position of the impact parameter -- the closest point between the sightline and the galactic center. (Bottom) The random distribution of impact parameters between 10 and 50 kpc. Note that the minimum impact parameter is 10 kpc, and the appearance of impact parameters $< 10$ kpc in the top panel is due to projection effects.}
\label{fig:sightline_description} 
\end{figure}

\begin{figure*}
\includegraphics[width=\textwidth]{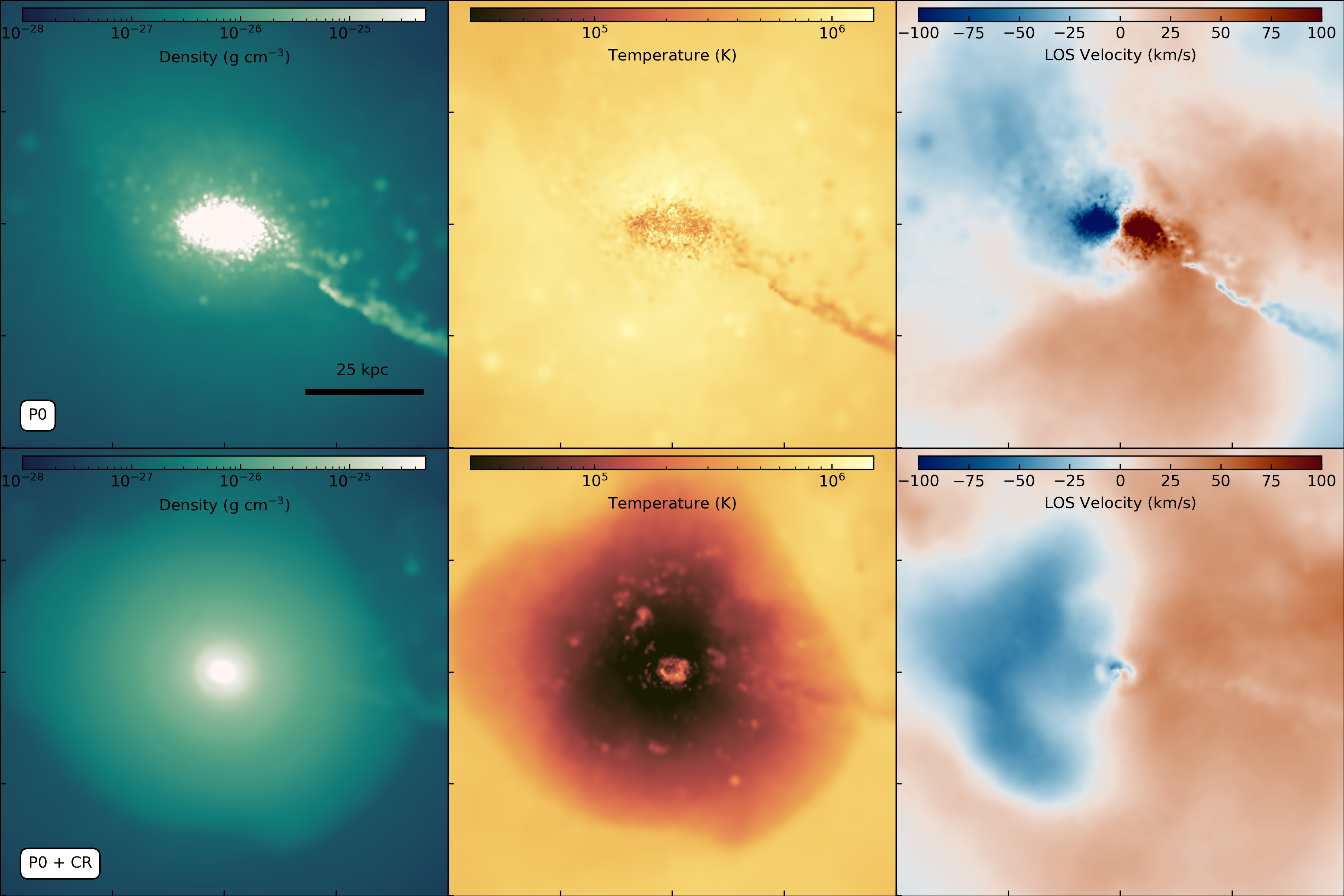}
\caption{ The mass-weighted projections of density, temperature and line-of-sight velocity for \po\ (top) and \pocr\ (bottom) at z = 0.25. The width and depth of the projection are 100 kpc. Relative to \po, the simulation with cosmic rays has a cooler CGM in the inner $\sim 40$ kpc and a larger rotating extended disk. }
\label{fig:multipanel1} 
\end{figure*}

\subsection{Generating Synthetic Spectra}
We explicitly track the evolution of hydrogen, oxygen, and iron at runtime. In the analysis phase, we use \trident \citep{Hummels:2017}, an extension of the widely-used simulation analysis tool, \yt \citep{Turk:2011}, to estimate the abundance of other chemical species as well as their ionization states. We also use \trident to generate synthetic spectra that mimic the specifications of the HST COS-G130M and COS-G160M instruments. 

We generate 100 randomly-oriented sightline coordinates that sample the inner CGM with impact parameters between 10 and 50 kpc. To generate the coordinates of a random sightline, we first define the position of the impact parameter -- the closest point between the sightline and the galactic center -- as a point on the surface of a sphere centered on the galaxy: $p(r, \theta, \phi)$. In this expression, $r$ is the impact parameter, $\theta$ is the polar angle, and $\phi$ is the azimuthal angle. All three values are selected using a random, unweighted number generator. Once we choose the coordinates of the impact parameter, we set the orientation of the sightline to be at a random angle, $\psi$, in the plane tangent to the sphere. All sightlines are 500 kpc long and centered at their impact parameter. 

We use the same set of galactocentric sightline coordinates to generate synthetic spectra for both, \po\ and \pocr. 
\autoref{fig:sightline_description} provides a visual depiction of the randomly distributed sightlines and the distribution of impact parameters (the circles). 

For each sightline in each galaxy, we generate a synthetic spectrum at $z = 0.25$, including all transitions of the  \hi, Mg~{\sc ii}, C~{\sc ii}, C~{\sc iii}, C~{\sc iv}, Si~{\sc ii}, Si~{\sc iii}, Si~{\sc iv}, N~{\sc v}, and \ovi\ ions. Each sightline starts as a 1D ray through the simulation. The gas properties along that ray (e.g., density, temperature, metallicity, velocity) are interpolated by \yt from the nearest gas particles within the smoothing length. Using these gas properties, \trident  generates Voigt absorption profiles that explicitly model Doppler broadening and collisional (pressure) broadening following the method described in \citet{Hummels:2017}. 
While \trident accounts for both thermal and turbulent pressures, it does not explicitly include cosmic-ray pressure when calculating line broadening. Unlike thermal Doppler broadening and nonthermal turbulent pressure broadening, cosmic-ray pressure does not directly contribute to the gas motions that broaden the Voigt profile. Instead, cosmic-ray pressure affects absorption line widths indirectly, by altering the density (and therefore the characteristic size) of cool gas clouds. Additionally, we do not include self-shielding, which may lead us to underpredict the column densities of \si\ and \ovi, especially at higher densities. However, we do not expect self-shielding to affect the ion kinematic profiles.  

When generating spectra, we subtract out the bulk velocity of the 500 kpc sphere centered on the galaxy relative to the simulation frame. To make these synthetic spectra resemble real HST/COS data, we add noise with a signal-to-noise ratio of 10 per resolution element and convolve it with the line-spread functions of the G130M and G160M spectrographs. The resulting normalized synthetic spectra are characterized by a FWHM $\approx$ 18 km s$^{-1}$ when they are binned by three native spectral pixels to a dispersion of $\Delta \lambda \approx$ 0.0367 \AA.

\subsection{Analyzing Synthetic Spectra}

Our spectral analysis incorporates common observational techniques for analysing medium-resolution HST/COS data \citep[e.g.,][]{Tumlinson:2011, Werk:2013}. Specifically, we extract information from the synthetic spectra by decomposing the absorption features into Voigt profiles. We do so without any regard to the parameters that are used by \trident to generate the synthetic spectra. 

The decomposition involves five steps: (1) automated identification of distinct absorption components; (2) an initial Voigt profile fit; (3) manual refinement of component definitions; (4) a final Voigt profile fit; and (5) calculation of column density upper limits for ions with no detected components. For each ion and sightline, this procedure produces either a column density upper limit or a collection of: component column densities, $N$;  velocity centroids relative to the halo rest frame, $v$;  and linewidths expressed as Doppler-$b$ parameters.

To identify absorption components for an ion along a sightline, we estimate the column density as a function of velocity, $N(v)$, and then split $N(v)$ into candidate components using a watershed-like segmentation algorithm. 
We use the apparent optical depth (AOD) method to estimate $N(v)$.
The AOD method is based on the assumption that the logarithm of the degree of absorption at a velocity, the apparent optical depth, is directly proportional to $N(v)$  \citep{Savage:1991}.
We then use the {\sc astrodendro} package \citep{Rosolowsky:2008} to split $N(v)$ up into a set of distinct and contiguous peaks; these are our candidate components. 
Finally, we remove candidate components whose integrated column densities are less than two times their integrated column density uncertainties.

Each candidate component has a centroid velocity, an allowed velocity range, and rough estimates of the component column density and linewidth.
These values are used as initial conditions for the Voigt profile fit.
We use optimization to find the likelihood-maximizing component parameter set and estimate uncertainties by calculating and inverting the Fisher information matrix \citep[e.g.,][]{Tegmark:1997} at the maximum likelihood solution.

The automated component identification procedure is imperfect.
It makes two main kinds of mistakes: (1) identifying a single component where multiple components would be more appropriate and (2) identifying a single absorption feature as arising from different ions.
We resolve the first kind of mistake by inspecting the initial Voigt profile fits and defining new components as needed.
The second mistake happens when ions have transitions with similar rest wavelengths, such as the C~{\sc ii}~1036\AA~and \ovi~1037\AA~lines.
In all such cases, at least one of the ions has more than one line, so the maximum likelihood solution for the fictitious component has a very low column density.
This allows us to find and remove these components using a simultaneous cut on velocity and column density.
A second round of Voigt profile fits done with the revised component structures provides our adopted set of component parameters.

If an ion has no components along a sightline, we calculate an upper limit on the ion's column density.
We use AOD to calculate $N(v)$ and integrate the result over a 100 km s$^{-1}$ interval. 
We then convert the total column density and column density uncertainty over this interval to a 2-$\sigma$ equivalent upper limit using the formalism of \citet{Bowen:2008}.

\section{Results}\label{sec:results}
\subsection{The Physical State of the CGM}
Despite starting from the same initial conditions, by a redshift of 0.25, the two galaxies have distinctly different CGM properties. While some differences are naturally to be expected (even for theoretically identical cosmological simulations) due to low-level non-linear effects \citep{Genel:2019}, the buildup of cosmic-ray pressure in \pocr\ produces a \textit{qualitatively} different CGM, significantly exceeding the differences expected from chaotic processes of galaxy formation. 

\autoref{fig:multipanel1} shows the mass-weighted projections of gas density, temperature, and line-of-sight velocity. The CGM of \po\ is notably hotter, with temperatures in excess of $10^6$ K, in part due to a recent burst of supernova feedback, and in part due to the thermal gas pressures required to maintain hydrostatic equilibrium. The cool, dense filament of inflowing gas traces the remnants of a recent merger with a satellite galaxy. 

By contrast, the inner CGM of \pocr\ is full of cool gas, supported against gravity and condensation by significant cosmic-ray pressure. This cool inner CGM extends roughly 40 kpc and has higher gas densities than the inner CGM of \po. The cool gas is pierced by pockets of hot bubbles from supernova feedback. There is a sharp transition between the cool, inner CGM and the outer CGM, marking the edge of the cosmic-ray-pressure-supported halo. The temperatures and densities in the outer CGM of \pocr\ resemble those in the outer CGM of \po. The line-of-sight velocity distribution is smoother in \pocr\ than in \po. The inflowing filament from the recent merger is also present in \pocr, but is significantly diminished relative to the fast-moving filament in \po.

\begin{figure}
\includegraphics[width=0.5\textwidth]{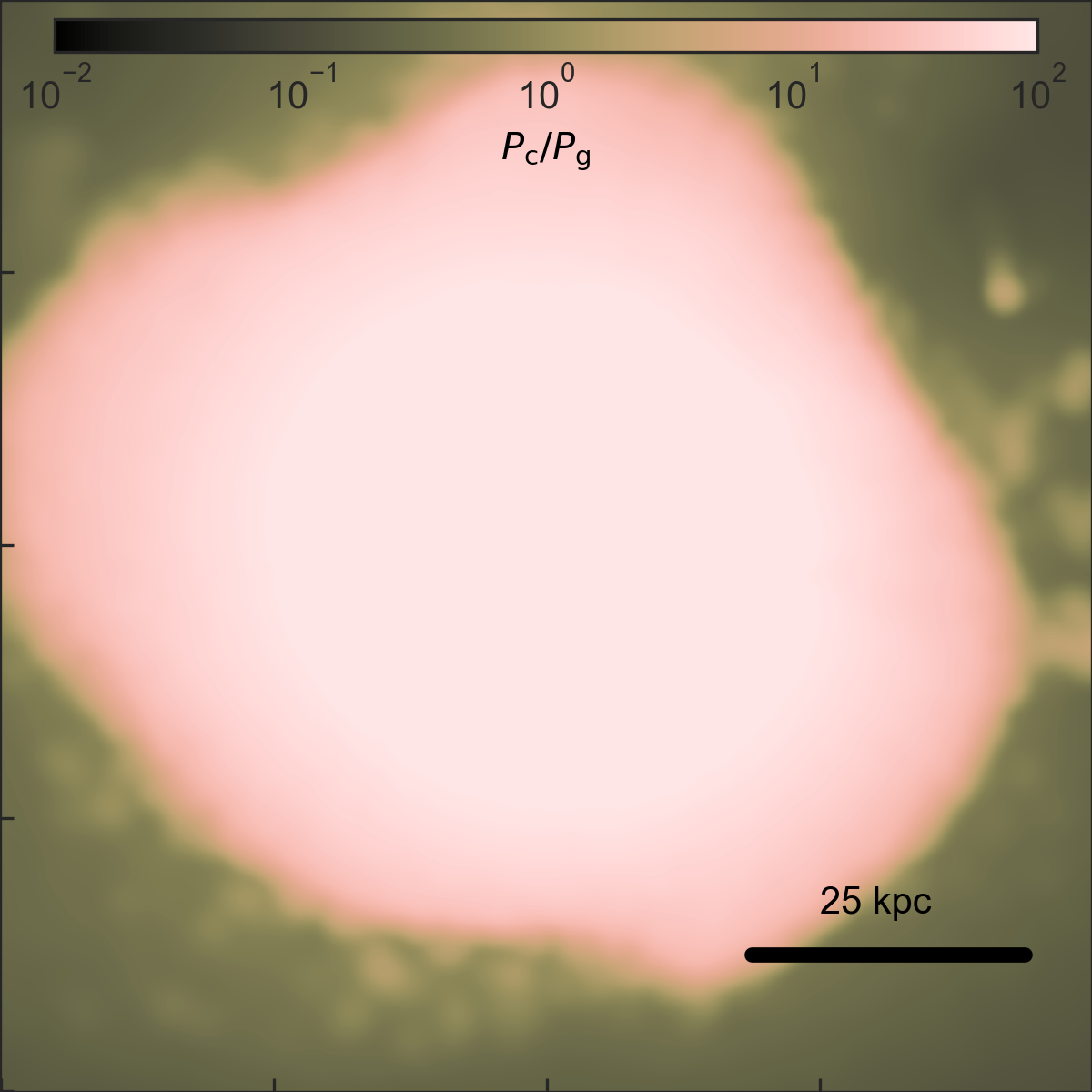}
\caption{   The mass-weighted projection of the ratio of cosmic-ray pressure to gas pressure in \pocr. Cosmic-ray pressure is significantly larger than thermal pressure in the inner CGM. The cosmic-ray pressure gradient counteracts gravity to maintain hydrostatic equilibrium, allowing cool, low-entropy gas to fill the inner CGM. }
\label{fig:cr_pressure} 
\end{figure}

\autoref{fig:cr_pressure} shows the projected ratio of cosmic-ray pressure to gas pressure around \pocr. The inner $\sim 40$ kpc of the CGM are supported by cosmic-ray pressure while the outer regions are supported by thermal pressure. The shape of the cosmic-ray-pressure-supported region traces the shape of the cool gas. The details of the cosmic-ray pressure profile and the extent of the cosmic-ray-pressure-supported region are sensitive to the choice of cosmic-ray hydrodynamics models, which are currently unconstrained. However, this simulation is qualitatively similar to the cosmic-ray-pressure-supported CGM seen in other hydrodynamics simulations with cosmic-ray diffusion \citep[e.g.,][]{Salem:2016, Butsky:2018, Ji:2020, Buck:2020}. 

\begin{figure*}
\includegraphics[width=\textwidth]{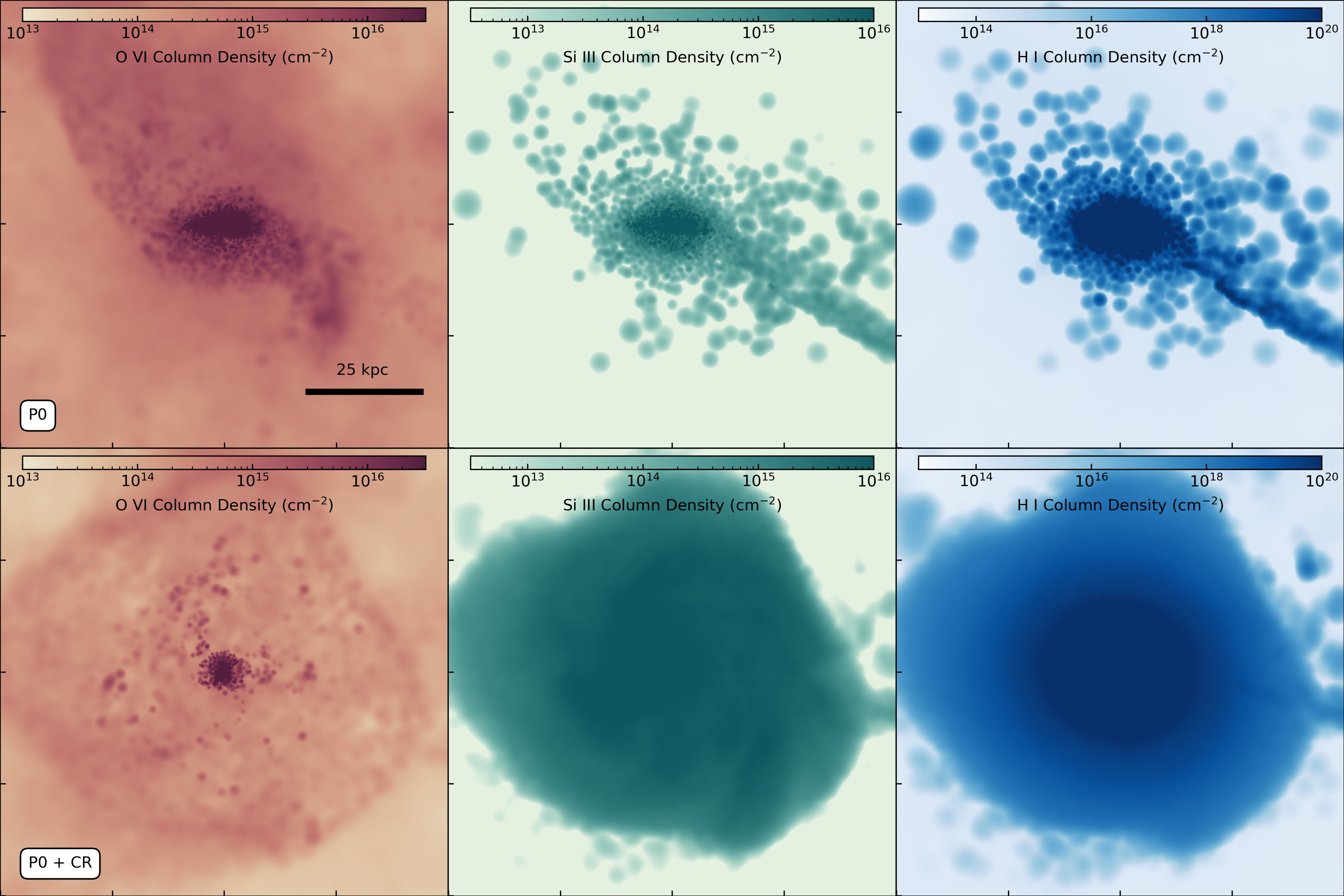}
\caption{   The \ovi, \si, and \hi\ column densities at z = 0.25 for \po\ (top) and \pocr\ (bottom). The width (and depth) of the projection is 100 kpc. Compared to \po, the CGM of \pocr\ has lower overall \ovi\ column densities and higher column densities of \si\ and \hi\ in the inner 50 kpc.}
\label{fig:multipanel2} 
\end{figure*}

These differences in the gas phase between the two galaxies translate into differences in their ion column densities. \autoref{fig:multipanel2} shows the column densities of \ovi, \si, and \hi\ for the two simulated galaxies. \po\ has high \ovi\ column densities throughout, with its hot inner region exceeding column densities of $10^{15} {\rm cm}^{-2}$. While the \ovi-bearing gas creates structures that span the extent of the inner CGM, the \si\ and \hi\ column density maps trace small-scale structures near the galactic disk and in the cold accretion stream. The sizes of the cloudlets that trace \si\ and \hi\ absorbers are likely limited by the resolution of the simulation \citep[e.g.,][]{Hummels:2019, Peeples:2019, Suresh:2019, vandeVoort:2019}

Relative to \po, \pocr\ has lower \ovi\ column densities with a profile that traces the shape of the cosmic-ray-pressure-supported region. The inner CGM has significant \si\ and \hi\ column densities throughout. 

\begin{figure*}
\includegraphics[width=0.5\textwidth]{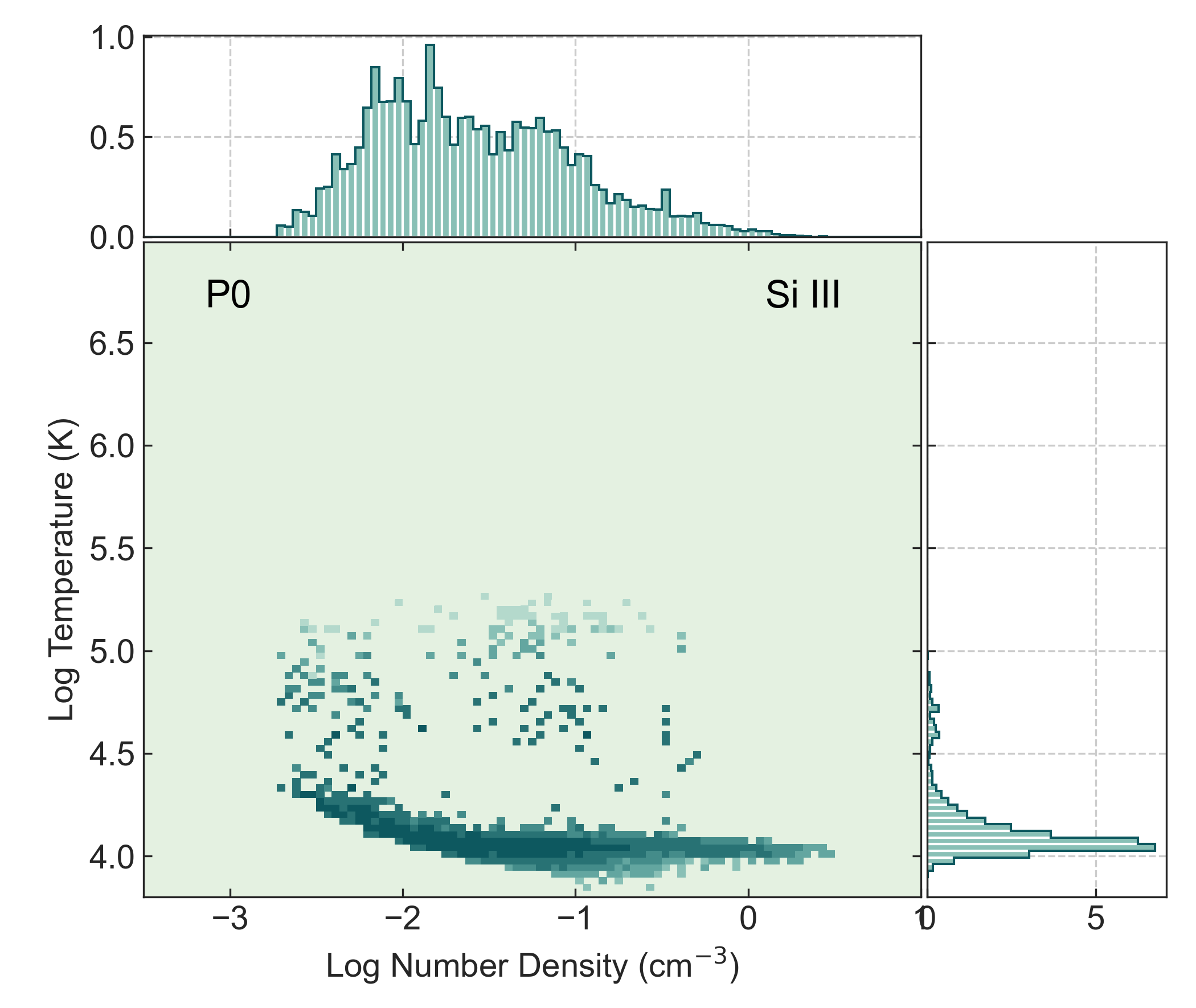}
\includegraphics[width=0.5\textwidth]{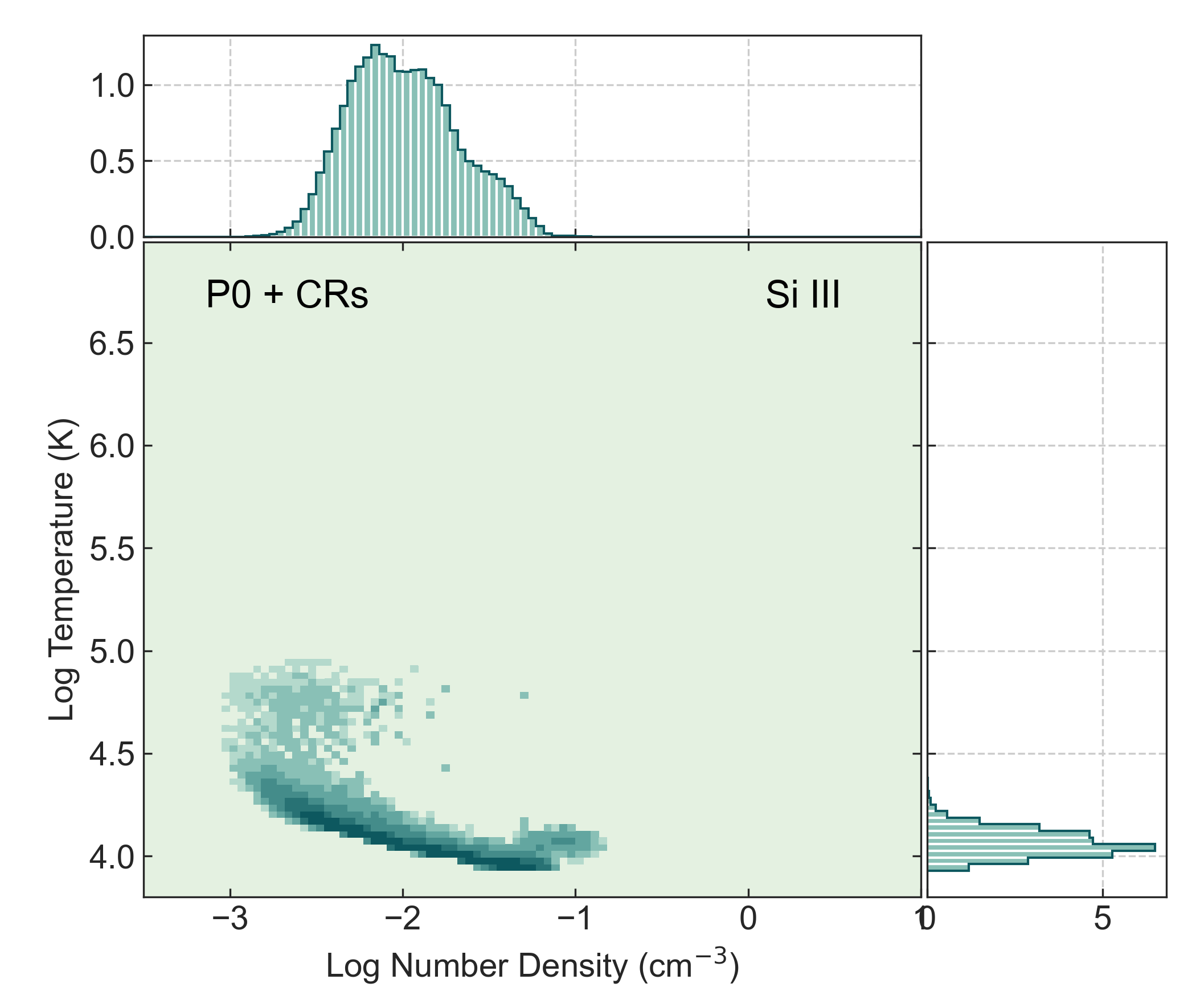}
\includegraphics[width=0.5\textwidth]{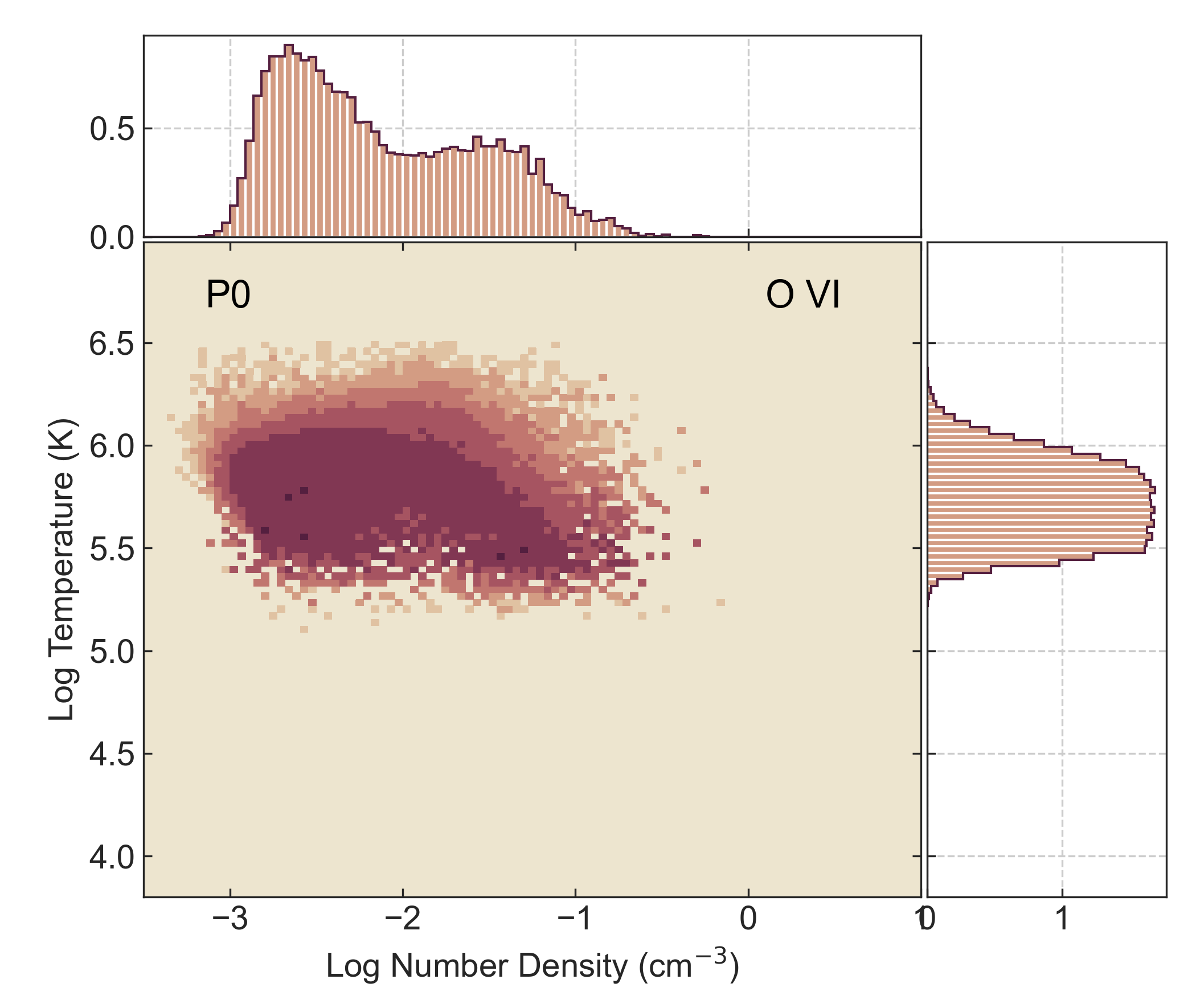}
\includegraphics[width=0.5\textwidth]{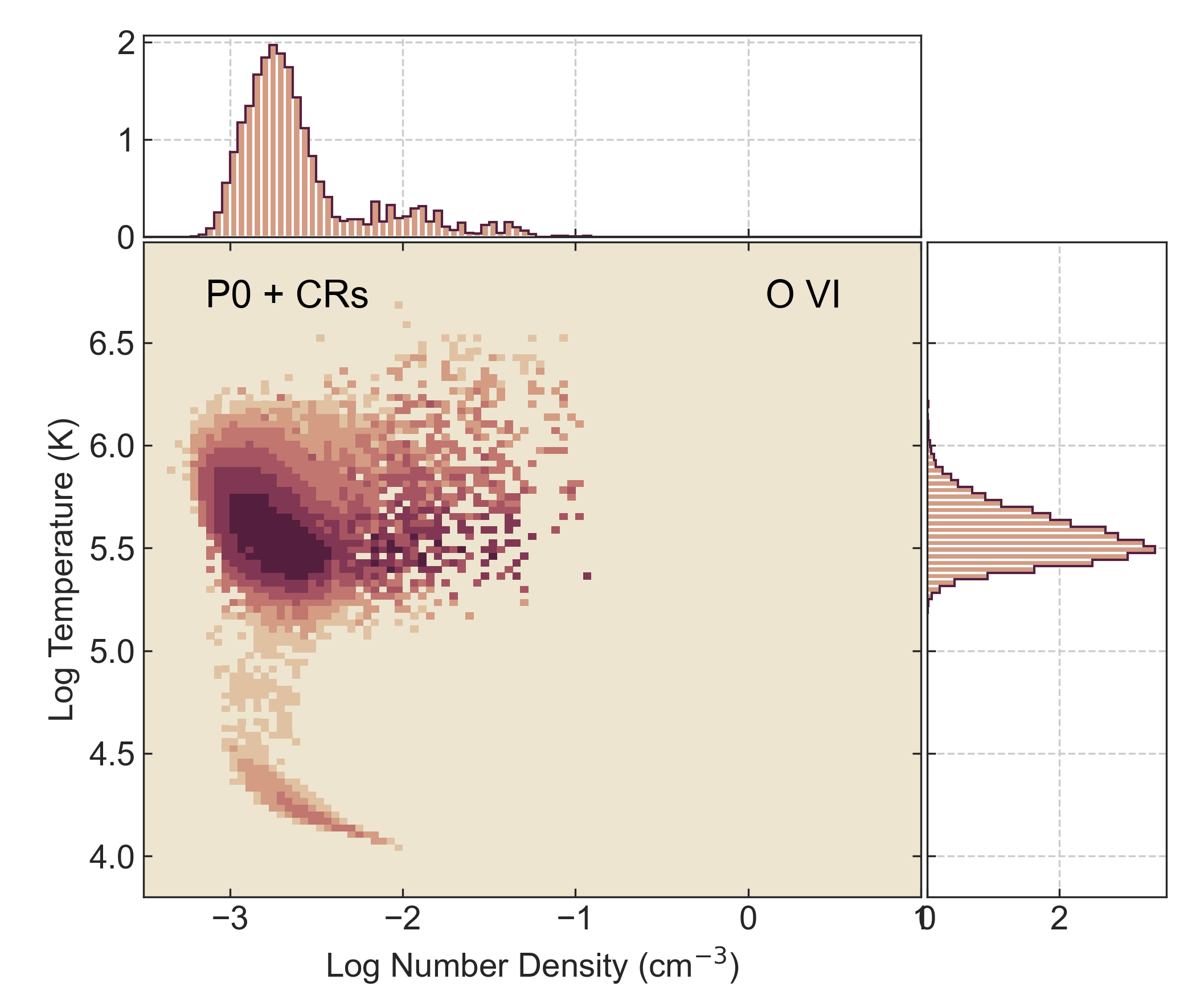}
\caption{   The ion-mass weighted 2D histograms of gas temperature and number density for \po\ (left) and \pocr\ (right). The data for these plots is generated from a spherical shell between 10 and 50 kpc from the galactic center. The additional pressure support from cosmic rays moves the gas to lower densities. \po\ has a substantial amount of \ovi\ absorption at temperatures above the collisional ionization equilibrium temperature of $T \sim 10^{5.5}$ K.}
\label{fig:phase} 
\end{figure*}

In \autoref{fig:phase}, we take a look at the differences in the gas properties of \ovi\ and \si\ absorbers between the two simulations. The images show the 2D histograms of simulated temperature and number density, colored by the ion mass probability density of \ovi\ and \si. The data for these plots was generated from spherical shells centered on the galaxy with radii between 10 and 50 kpc, corresponding to the region sampled with our synthetic spectra and in the previous figures.

In both galaxies, the vast majority of \si\ comes from gas around $10^{4.2}$ K, as is expected for photoionized gas. Relative to \po, the \si\ in \pocr\ comes from lower density gas due to nonthermal cosmic-ray pressure support. 
In \po, \ovi\ traces gas with temperatures of $T = 10^{5.3 - 6.2}$ K --- hotter than the peak \ovi\ ionization temperature of $\sim 10^{5.5}$ K. In \pocr, the majority of \ovi\ absorption traces gas with temperatures with a peak at $T = 10^{5.5}$ K and has a negligible fraction of photoionized \ovi\ at low temperatures and low gas densities. Like with \si, \ovi-bearing gas in \pocr\ tends to have lower densities than in \po. 

\begin{figure*}
\includegraphics[width=0.5\textwidth]{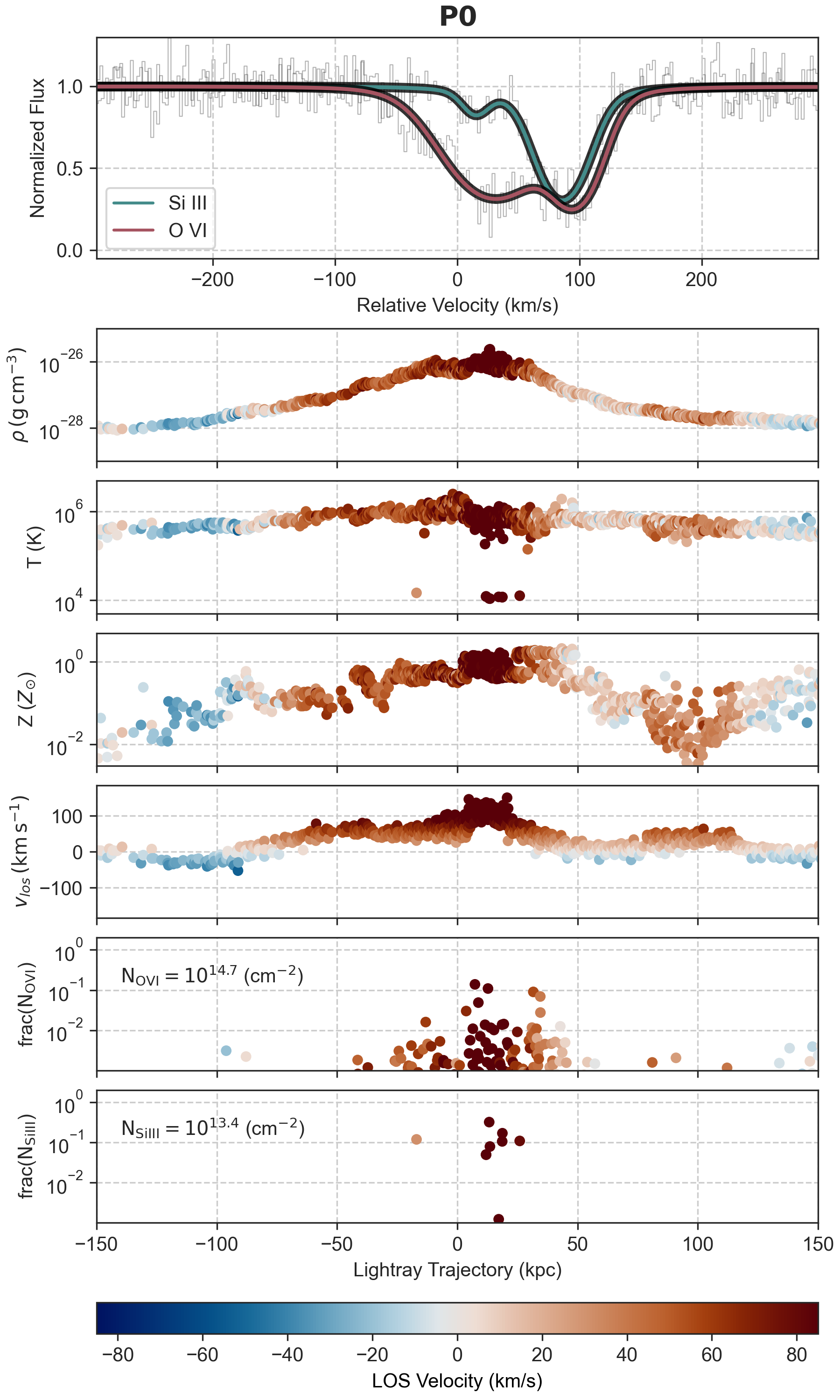}
\includegraphics[width=0.5\textwidth]{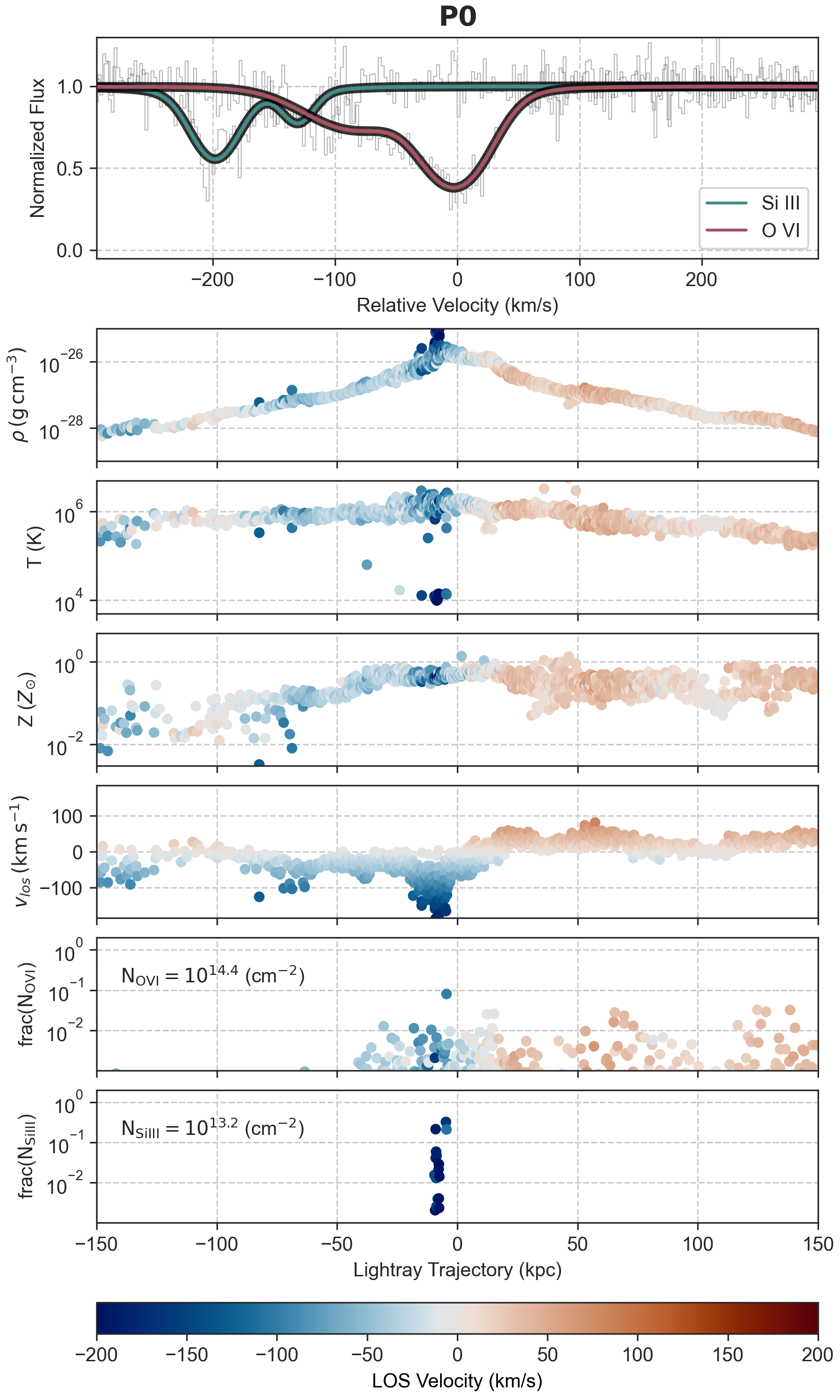}

\caption{Example synthetic spectra from \po\ and the gas properties along their line of sight. (Top row) The normalized flux corresponding to \si\ and \ovi\ absorbers in velocity space. The gray lines show the raw synthetic spectra and the green and red lines show the Voigt profile fits for \si\ and \ovi\ respectively. (Remaining rows) The gas density, temperature, metallicity, line-of-sight velocity, and fractional ion column densities as a function of position along the line of sight. The points are colored by their line-of-sight velocity, which helps discern which absorber they contribute to. The fractional ion column densities in the bottom row are divided by the total ion column density, to help discern which particles contributed the most to the absorption features in the top row. In the light ray trajectory, 0 always represents the impact parameter (point of closest approach to the galaxy center). The impact parameters of the sightlines in the left and right panels are 21.1 kpc and 11.3 kpc respectively. While \ovi-bearing gas tends to form large, resolved structures, \si-bearing gas tends to form dense, resolution-limited structures. By matching fractional column densities of \si\ and \ovi\ to the absorbers in the spectra, we can see that kinematic alignment does not necessarily imply physical alignment. \po\ spectra have many cases of either kinematic misalignment between \si\ and \ovi, or kinematic alignment in which there is no physical correlation between the absorbers. Conversely, the right panel shows an example of \si\ and \ovi\ absorbers that are spatially aligned but kinematically misaligned.}

\label{fig:spectra} 
\end{figure*}

\begin{figure*}
\includegraphics[width=0.5\textwidth]{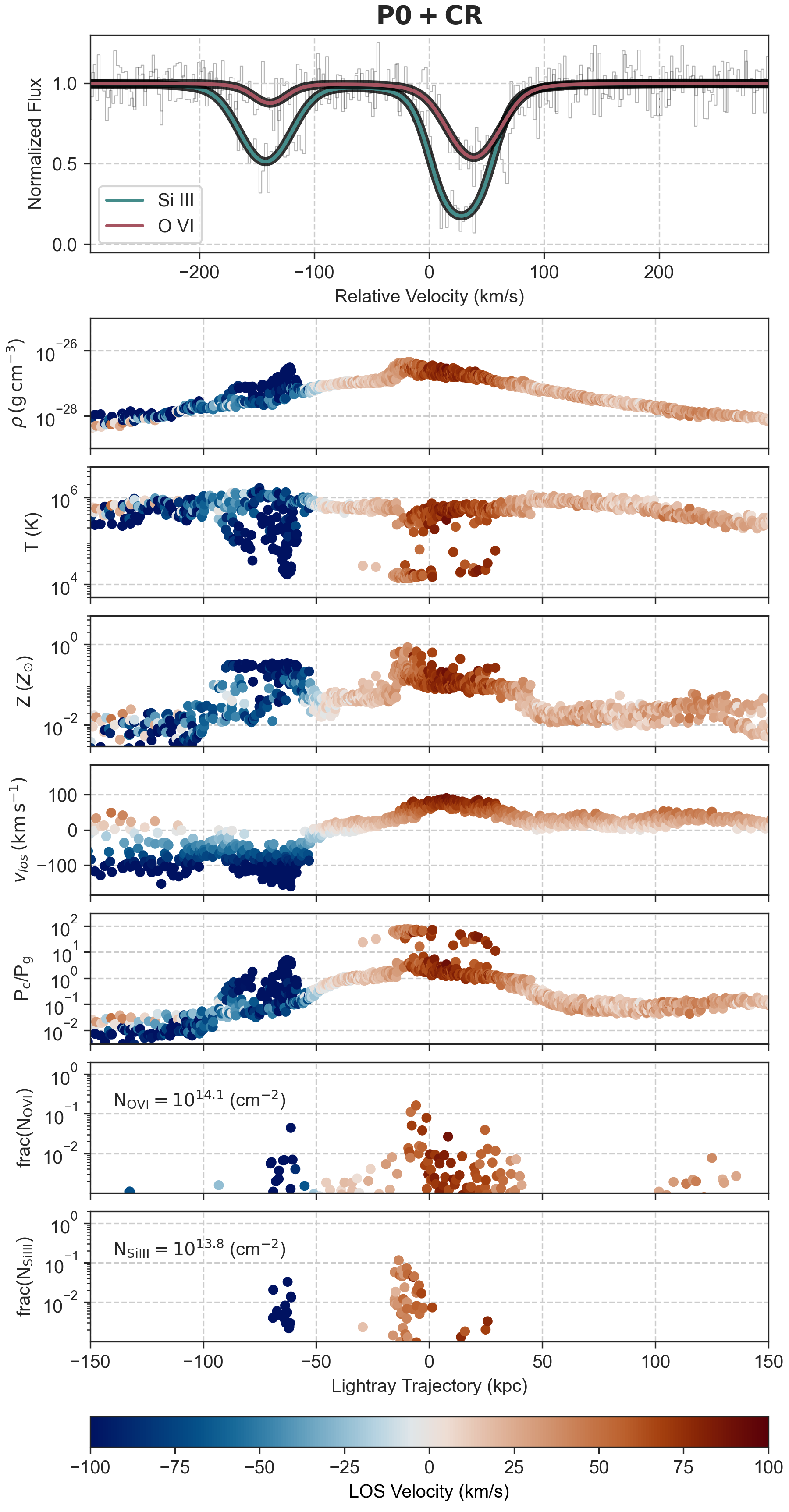}
\includegraphics[width=0.5\textwidth]{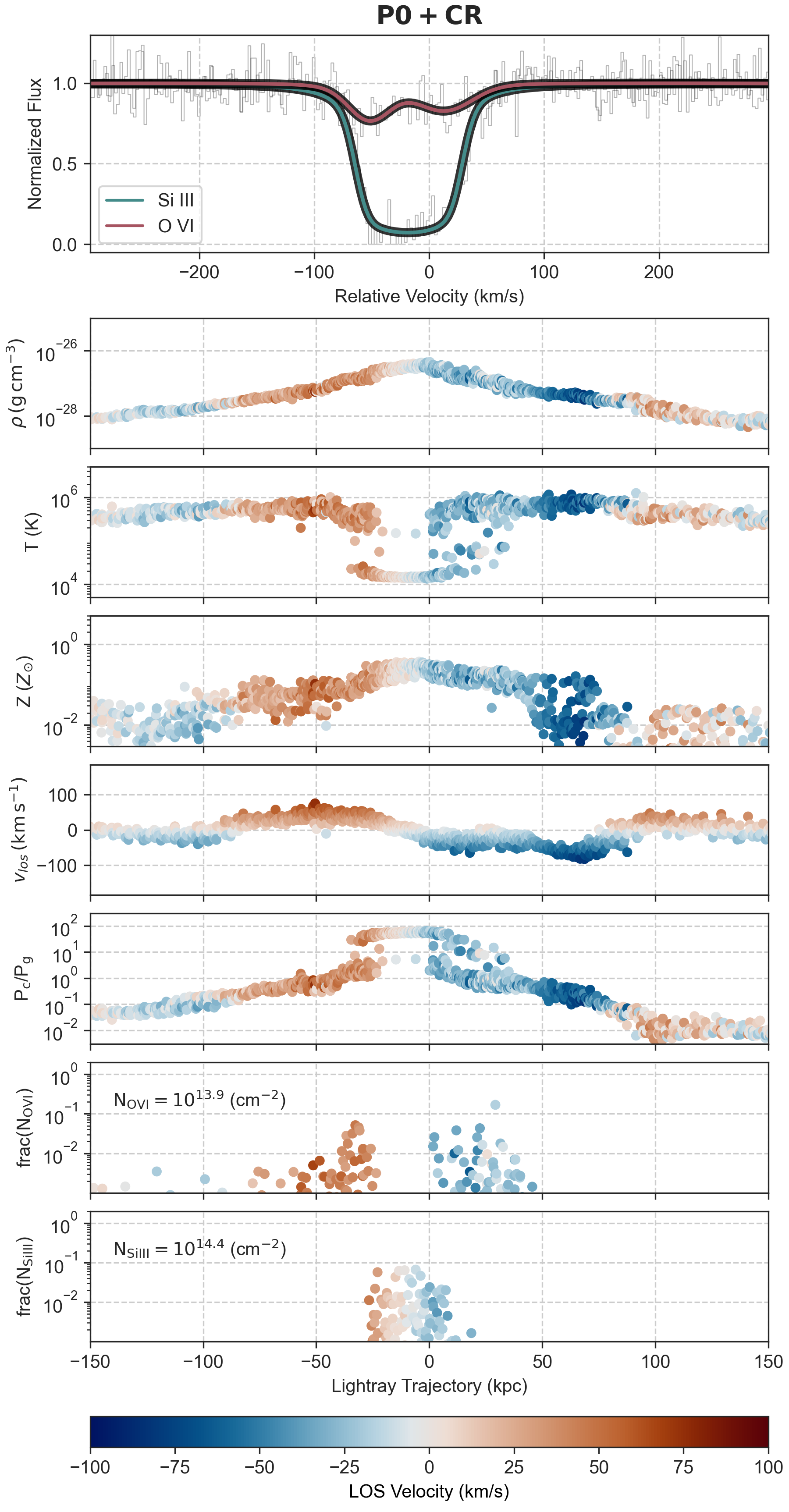}
\caption{ Similar to \autoref{fig:spectra}, the figure above shows example synthetic spectra from \pocr\ and the gas properties along their line of sight. There is an additional panel to show the ratio of cosmic-ray pressure to gas pressure, $\rm P_{\rm c}/P_{\rm g}$, along the line of sight. The impact parameters of the sightlines in the left and right panels are 36.5 kpc and 43.0 kpc respectively. (Left) An example of kinematic alignment in \pocr\ in which both \si\ and \ovi\ absorbers come from the same physical structures. In all \pocr\ spectra that have both \ovi\ and \si\ absorbers, there is at least one \ovi\ component that is kinematically aligned with \si. \pocr\ has higher incidences of kinematic alignment between \si\ and \ovi. (Right) In most cases in \pocr, \si\ absorption comes from large, $\sim 10-50$ kpc, clouds that are enveloped by \ovi\ absorbers. Unlike in \po, low-density \si-bearing gas in \pocr\ is resolved by many gas particles. }
\label{fig:spectra_cr} 
\end{figure*}

\begin{figure*}
\includegraphics[width=0.5\textwidth]{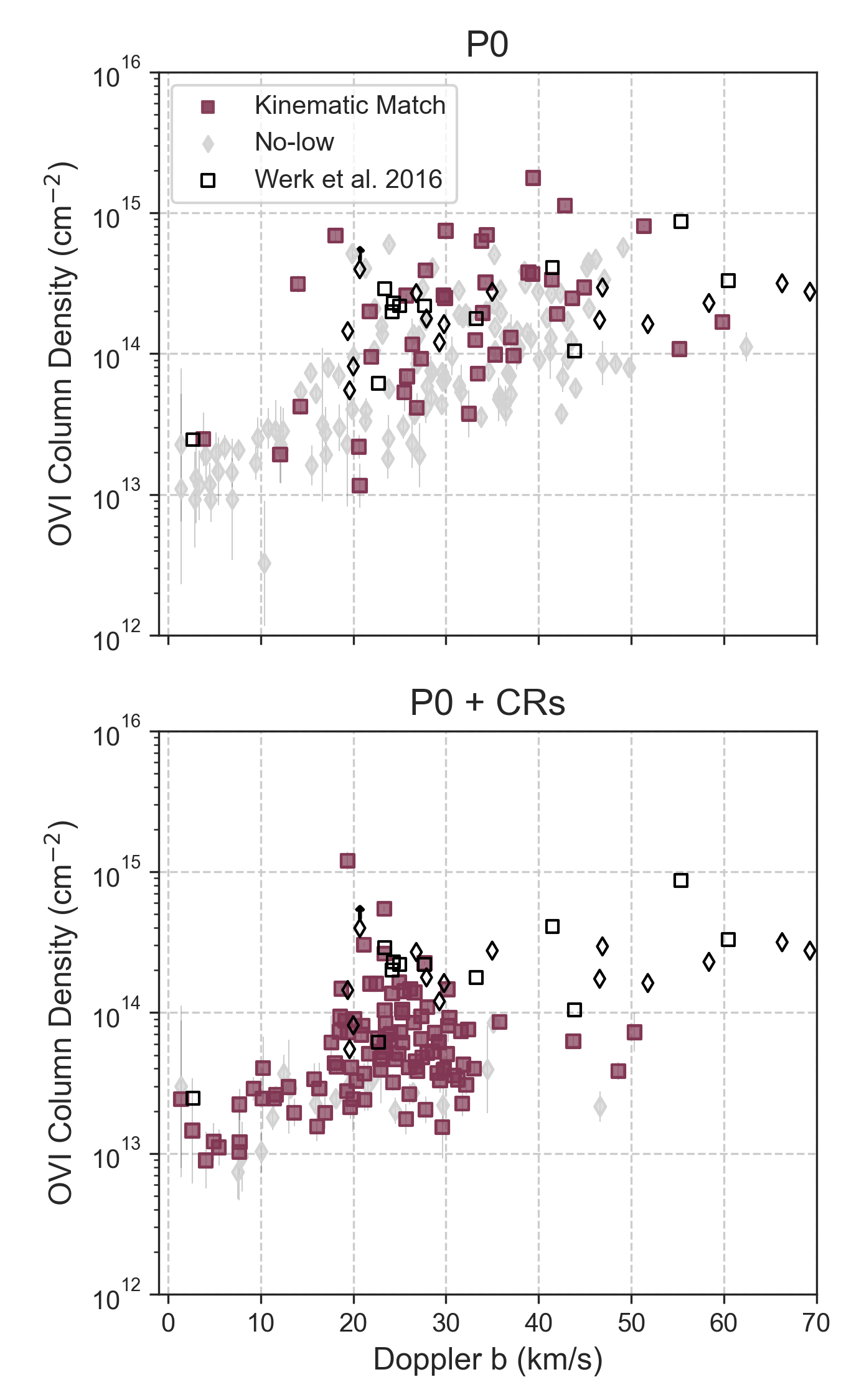}
\includegraphics[width=0.5\textwidth]{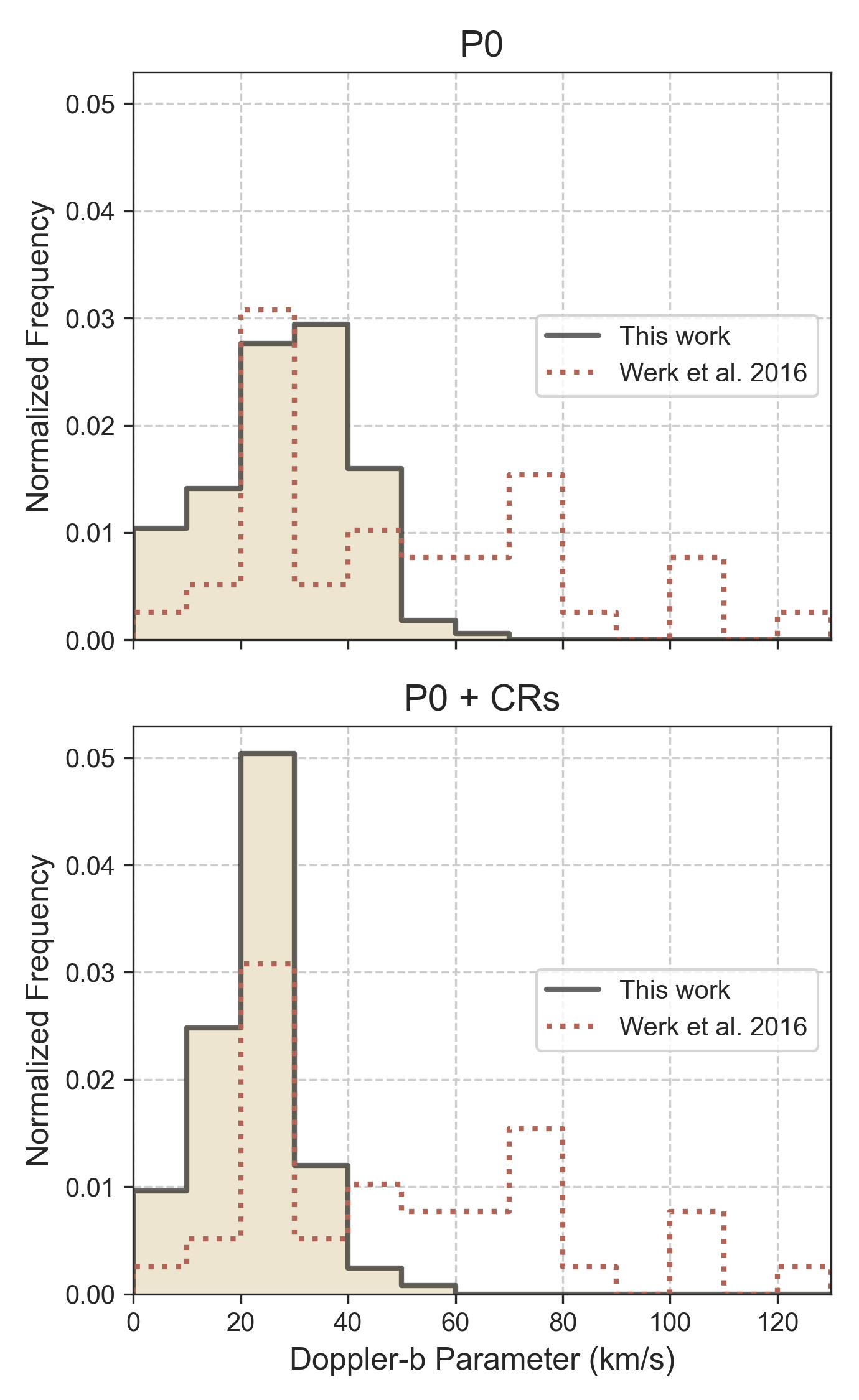}
\caption{ (Left) The column density of \ovi\ absorbers as a function of Doppler-$b$ parameter for \po\ (top) and \pocr\ (bottom). Purple squares indicate \ovi\ absorbers that are matched with a \si\ absorber within 35 km/s in line-of-sight velocity space. Gray diamonds indicate \ovi\ absorbers that have no corresponding \si\ absorber within 35 km/s. We include data from \citet{Werk:2016} for comparison in black. Upward arrows indicate upper limits on the measured column density. (Right) The histograms of the Doppler-$b$ parameters of \ovi\ absorbers in \po\ (top) and \pocr\ (bottom) compared against observations. All histograms are normalized to have an area of 1. \pocr predicts a relative abundance of narrow ($b = 20-30$ km/s) \ovi\ absorbers. Both simulations have a distinct lack of very broad ($b > 60$km/s) \ovi\ absorbers. }
\label{fig:ovi_bval} 
\end{figure*}

\begin{figure*}
\includegraphics[width=0.5\textwidth]{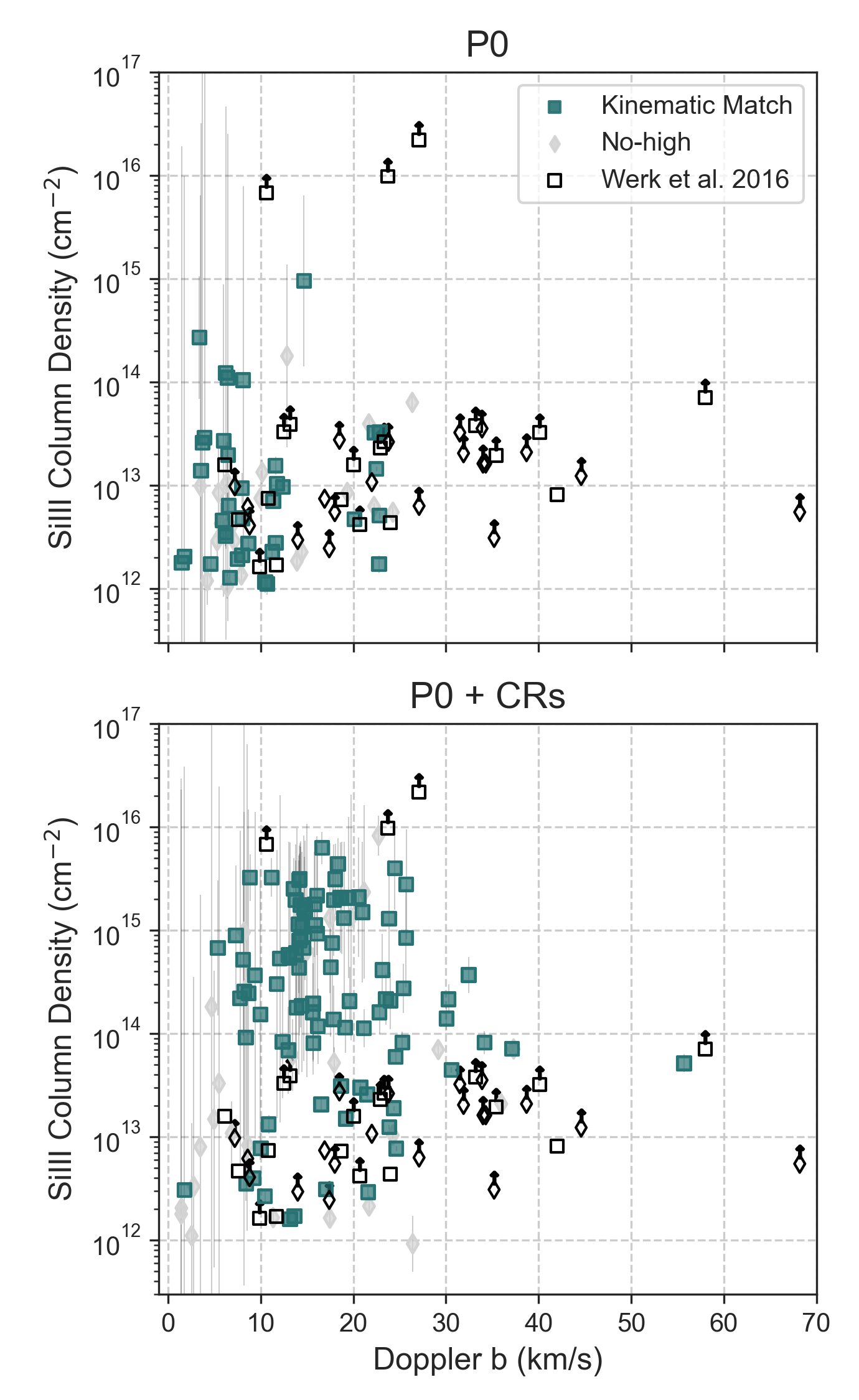}
\includegraphics[width=0.5\textwidth]{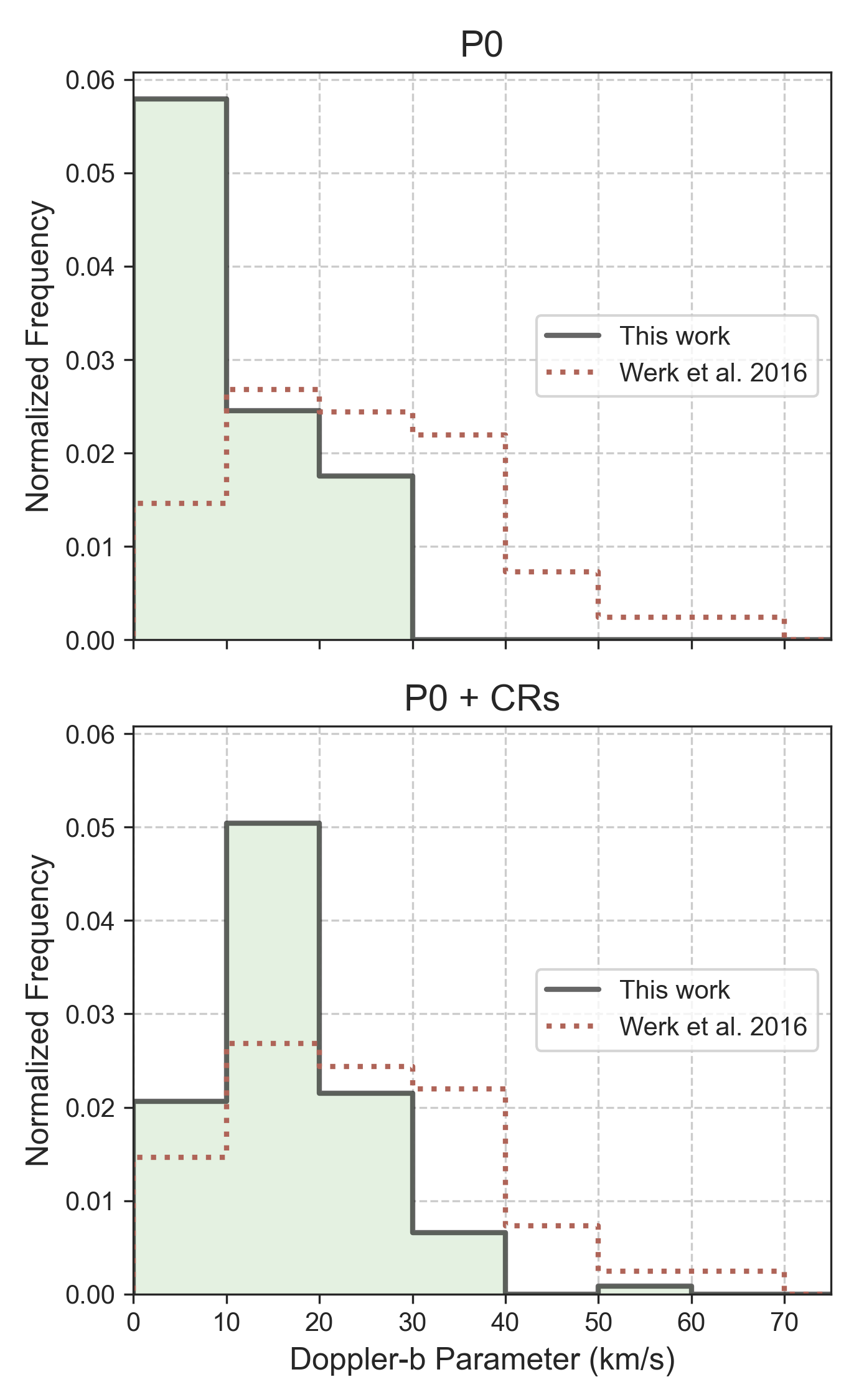}
\caption{(Left) The column density of \si\ absorbers as a function of Doppler-$b$ parameter for \po\ (top) and \pocr\ (bottom).
Green squares indicate \si\ absorbers that are matched with a \ovi\ absorber within 35 km/s in line-of-sight velocity space. 
Gray diamonds indicate \si\ absorbers that have no corresponding \ovi\ absorber within 35 km/s. We include data from \citet{Werk:2016} for comparison in black. Upward arrows indicate upper limits on the measured column density. (Right) The histograms of the Doppler-$b$ parameters of \si\ absorbers in \po\ (top) and \pocr\ (bottom) compared against observations. All histograms are normalized to have an area of 1. \pocr\ better reproduces the observed broad \si\ absorbers but over-predicts the abundance of narrow, high-column \si\ absorbers. }
\label{fig:siiii_bval} 
\end{figure*}

\begin{figure}
\includegraphics[width=0.5\textwidth]{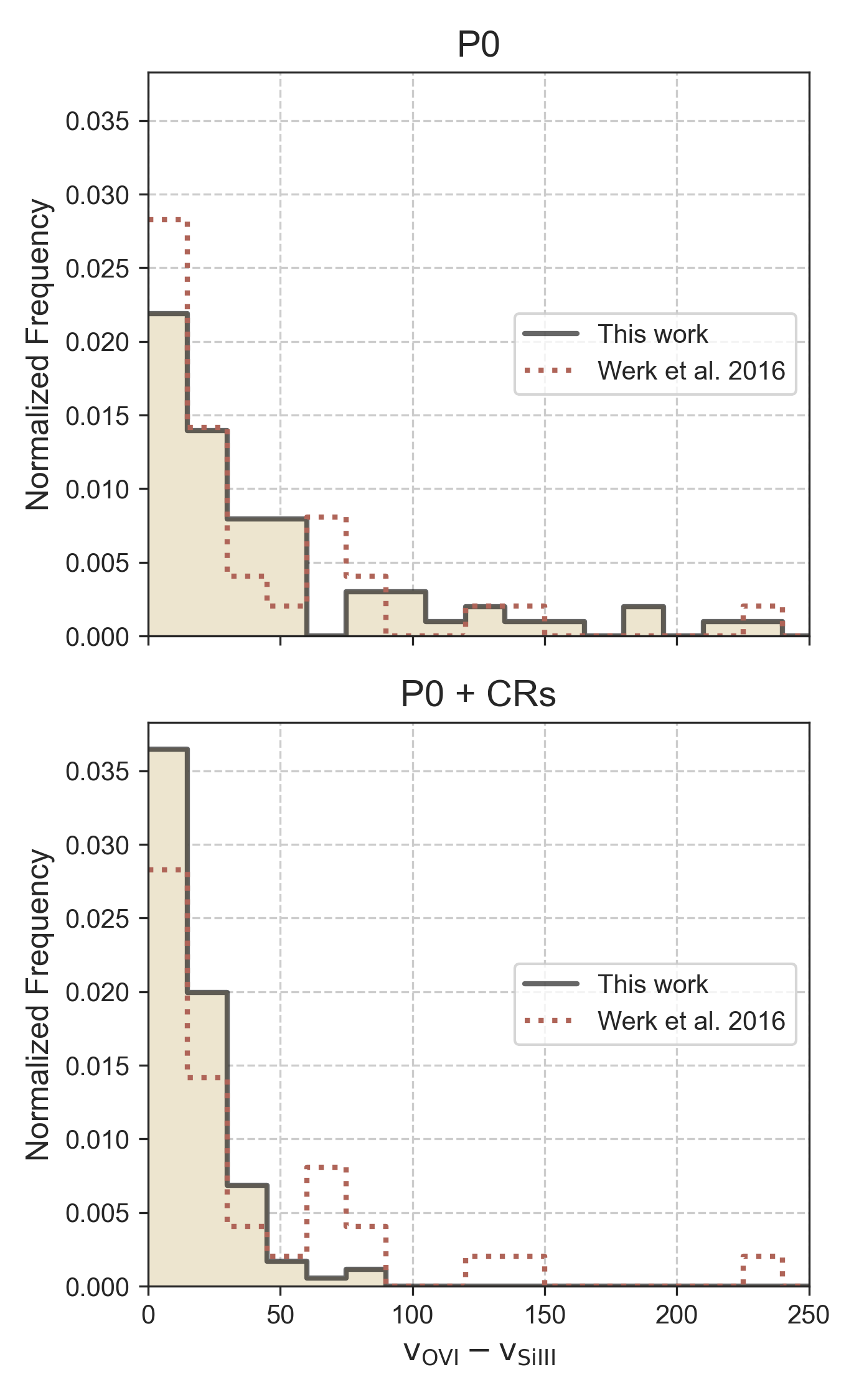}
\caption{ The histograms of the minimum velocity differences between \ovi\ and \si\ absorbers for \po\ (top) and \pocr\ (bottom). This excludes sightlines that have \ovi\ but no \si\ absorbers. All histograms are normalized to have an area of 1. \pocr\ better reproduces the observed kinematic link between \ovi\ and \si\ absorbers. }
\label{fig:ovi_dvel} 
\end{figure}

\begin{figure}
\includegraphics[width=0.5\textwidth]{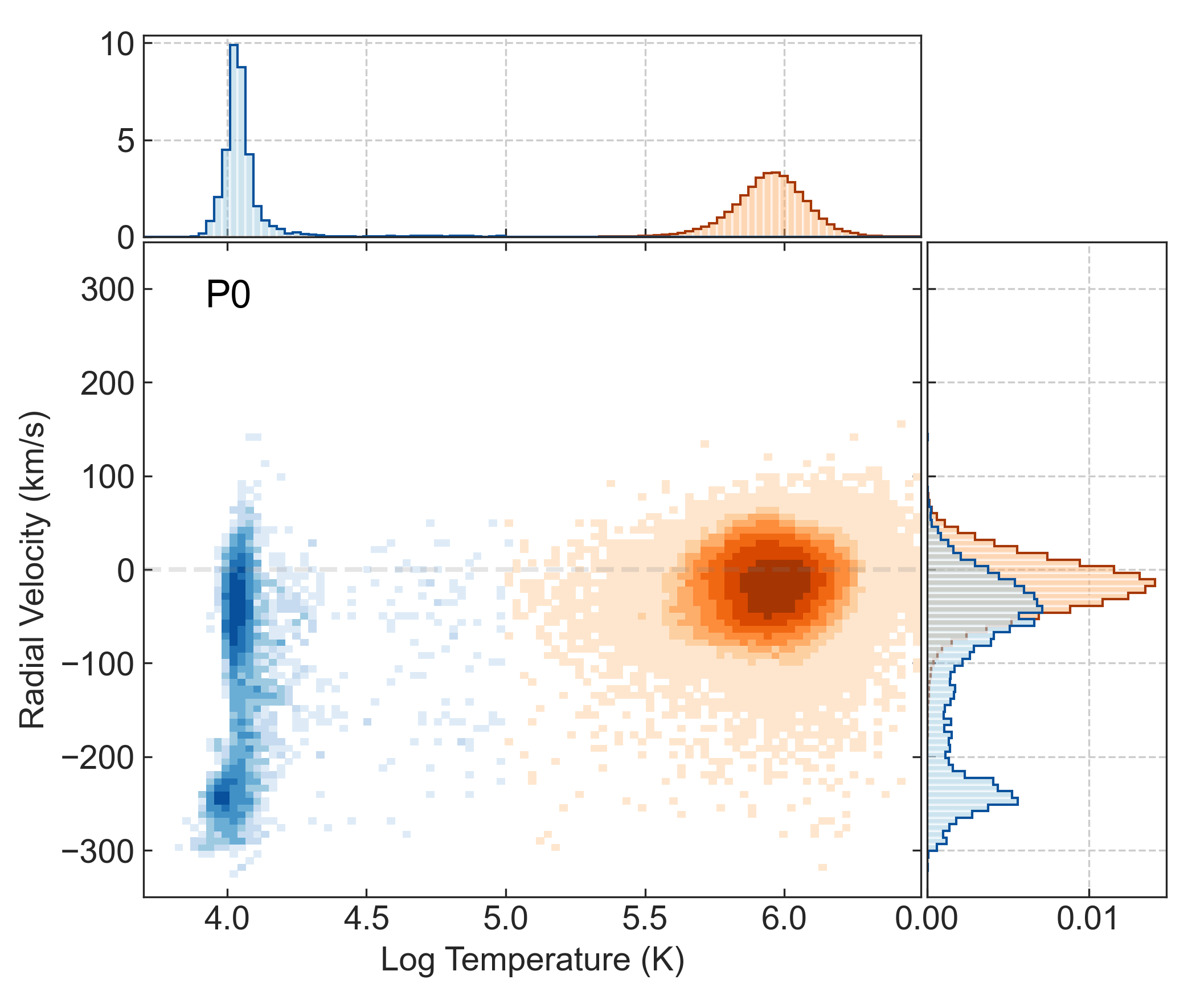}\\
\includegraphics[width=0.5\textwidth]{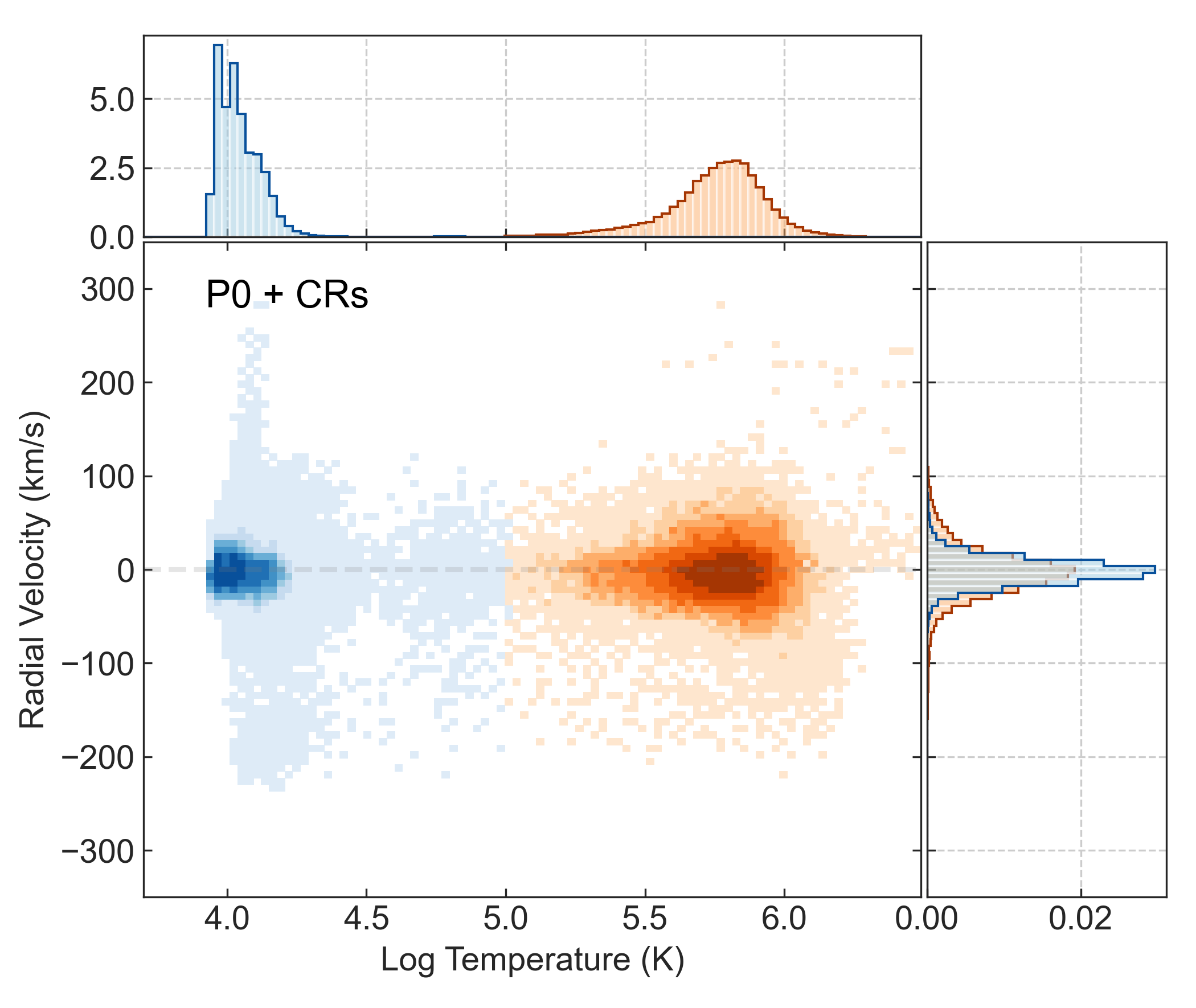}
\caption{ The mass-weighted 2D histogram of radial velocity as a function of temperature for cool (blue) and warm (orange) CGM gas for \po\ (left) and \pocr\ (right). The data is taken from gas that is between 10 and 50 kpc from the galactic center. Without cosmic rays, cool gas in the inner CGM tends to be on a radially inward trajectory. With substantial cosmic-ray pressure support, low-entropy cool gas is supported against gravity and has a very similar velocity distribution to that of the warm gas, leading to kinematic alignment of multiphase absorbers. }
\label{fig:phase_temperature} 
\end{figure}

\begin{figure*}
\includegraphics[width=\textwidth]{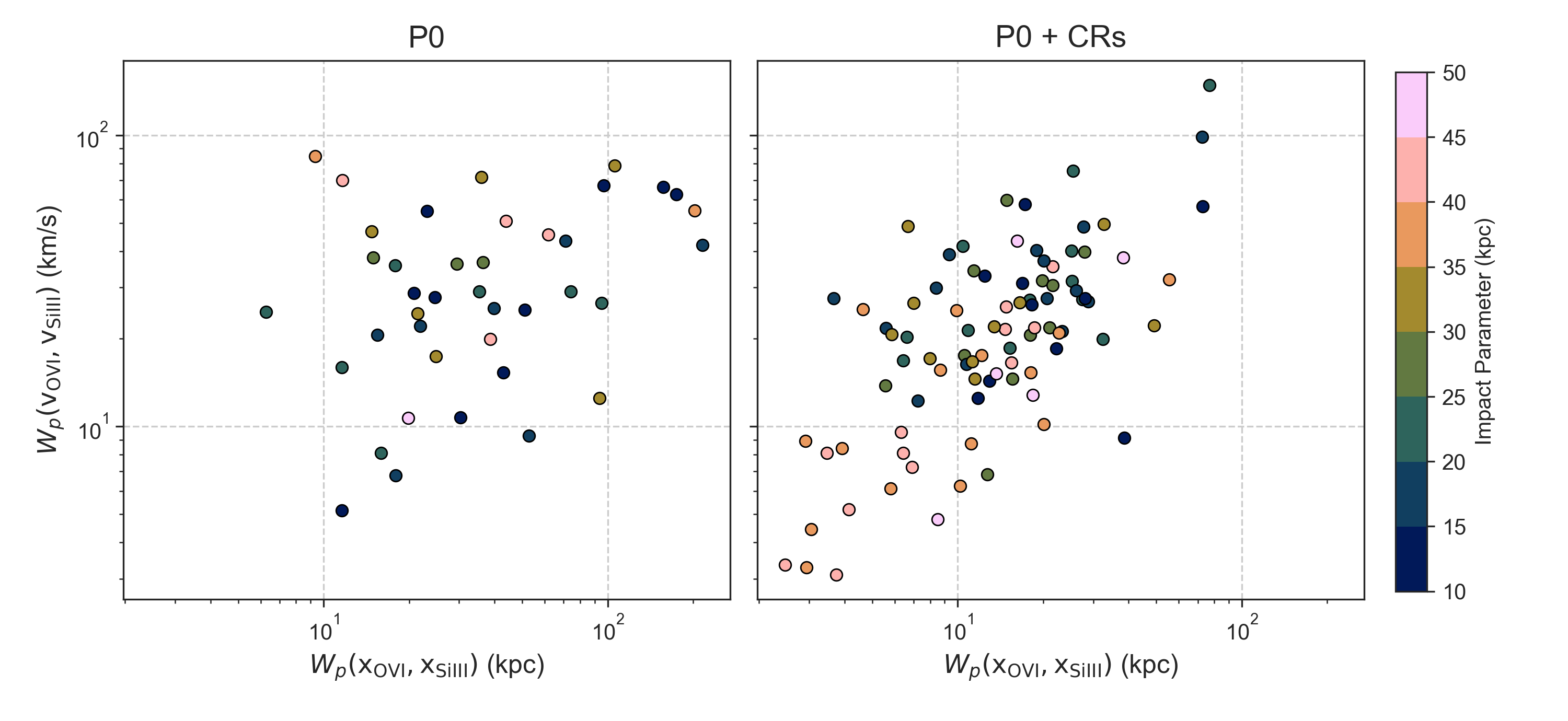}
\caption{ The Wasserstein distance between \ovi\ and \si\ line-of-sight velocities and positions along the sightline. The color of the points indicates the impact parameter of the sightline.
\pocr\ has a clear relationship between the kinematic and spatial alignment of \ovi\ and \si\ absorbing gas. In contrast, \ovi\ and \si\ gas in \po\ are less likely to be kinematically or spatially aligned. }
\label{fig:wass} 
\end{figure*}
%%%%%%%%%%%%%%%%%%%%%%%%%% Synthetic Spectra %%%%%%%%%%%%%%%%%%%%%%%%%%%%%%%%%
\subsection{Physical Insights from Synthetic Spectra}
Next, we present the physical state of the two galaxies as seen through their synthetic spectra. Unlike real observations, synthetic spectra have the additional benefit of linking absorption features to the exact gas properties that gave rise to them. \autoref{fig:spectra} and \autoref{fig:spectra_cr} show two examples of synthetic spectra that detect both \si\ and \ovi. The top panels show the ion absorption as a function of its velocity offset from the galaxy rest frame. The remaining panels show various properties of the gas particles along the one-dimensional sightline used to generate the synthetic spectra. Unlike the top panel, the x-axis of the remaining panels shows the spatial trajectory along the sightline. In all cases, the sightline trajectories are centered so that zero corresponds to the location of the impact parameter. The scattered points are colored by their line-of-sight velocity, which provides a visual aid to determine which gas particles contributed to which absorption features in the top panel. 

Figure \ref{fig:spectra} highlights two example synthetic spectra from \po. 
The \po\ spectrum on the left contains two different examples of kinematic alignment between \ovi\ and \si. The fast-moving \ovi\ absorber, with a relative velocity of $\sim 100$ km/s comes from the same physical region as the correlated \si\ absorber (see the spatial alignment of the deep-red points in the bottom two panels of \autoref{fig:spectra}). 
This is consistent with cold clumps being entrained in a hot background medium. 
By contrast, the \ovi\ and \si\ absorbers moving with a relative velocity of $\sim20$ km/s come from completely different regions, separated by roughly 50 kpc along the line of sight. In this case, their kinematic alignment appears to be a coincidence. Additionally, the 20 km/s \si\ absorber comes from one extrapolated data point along the sightline (at $\sim -15$ kpc), indicating that the cool cloud is not resolved and resulting in a narrow \si\ absorption feature. 

The second \po\ spectrum, in the right panel of \autoref{fig:spectra}, contrasts the first example in that it contains two absorption features of both \ovi\ and \si\ without any kinematic alignment. The fast-moving \si\ absorber with a relative velocity of $-200$ km/s traces the inflowing accretion of cool gas following a recent merger. Although it is spatially aligned with some \ovi-absorbing gas, the line-of-sight velocity of this \si\ absorber is significantly offset from any \ovi\ features. The \si\ absorber with a relative velocity of $-120$ km/s appears to come from a single particle (medium blue point at $\sim -5$ kpc). Although that particle is spatially aligned with an \ovi\ cloud with similar line-of-sight velocities, the corresponding absorption features do not appear aligned. This is likely because the \ovi\ absorption feature is broadened by gas particles moving with similar, but slightly different velocities between -50 and -100 km/s, whereas the unresolved \si\ absorber is only composed of one particle. 

The fact that the \si\ structures are unresolved likely contributes to artificially narrow \si\ absorbers, because the velocity structure of a fully-resolved cloud would contribute to line broadening \citep{Peeples:2019}.

The \pocr\ spectra in \autoref{fig:spectra_cr} show a very different relationship between the kinematic and physical alignment of \ovi\ and \si. In both example spectra, the \ovi\ and \si\ absorbers are kinematically aligned. 
The first example spectrum, in the left panel of \autoref{fig:spectra_cr}, has two distinct absorbers of each \si\ and \ovi. Both sets of kinematically aligned absorbers are also spatially aligned along the line of sight.

In the second example spectrum, in the right panel of \autoref{fig:spectra_cr}, the cool and warm gas comes from physically distinct, yet related clouds. Although \ovi\ and \si\ absorption is spatially offset, the \ovi\ absorption borders the \si\ absorption on either side. The velocity structure of the scattered points clearly shows that together, the \ovi\ and \si\ form a coherent physical region in which the \ovi\ and the \si\ absorbers are co-rotating. The \si\ absorption line is very broad because it comes from a large \si-bearing cloud that spans a wide range of velocity space. The two \ovi\ absorption features, which trace the velocity at either edge of the cool cloud are nestled within the broad \si\ absorber. 

These four examples highlight the observed trends in and differences between the \po\ and \pocr\ synthetic spectra. \footnote{The interested reader can see this plot for every spectrum in our sample at \href{irynabutsky.me/data}{irynabutsky.me/data}}  
\po\ tends to have broader \ovi\ absorbers and more narrow \si\ absorbers. A large fraction of spectra either have no \si\ absorption or have \si\ absorbers that are kinematically misaligned with \ovi\ absorbers. \ovi\ absorbers often come from hot ($T > 10^6$ K) gas that spans tens of kpc while \si\ absorbers come from small (and likely resolution-limited) clouds of $\sim 1$ kpc scales. The \ovi- and \si- bearing gas sometimes comes from the same physical cloud, but is often physically unrelated. 

In \pocr, \si\ absorbers tend to be broader and come from large, low-density clouds spanning $10-50$ kpc. \ovi\ absorbers tend to come from slightly cooler gas ($T \lesssim 10^6$ K) and slightly smaller clouds ($10-50$ kpc) than in \po. \ovi\ and \si\ absorbers are often  kinematically aligned and often come from either the same physical cloud or from physically related clouds. 

Note that in an attempt to highlight the trends in the kinematic and physical relationships between \ovi\ and \si, we omit spectra that have no measurable \si\ absorption from \autoref{fig:spectra} and \autoref{fig:spectra_cr}. While these true ``no-lows" \citep[\ovi\ absorbers without a kinematically matched \si\ absorber; ][]{Werk:2016} are present in both simulations, they are very common in \po\ and very rare in \pocr. 

%%%%%%%%%%%%%%%%%%%%%%%%%% Doppler-b %%%%%%%%%%%%%%%%%%%%%%%%%%%%%%%%%
\subsection{The Doppler-b Parameter}
The Doppler-$b$ parameter is a measure of both thermal and nonthermal line broadening mechanisms, $b = \sqrt{b_{\rm th} + b_{\rm nt}}$.
The Voigt profile of the synthetic spectra generated with {\sc trident} is a convolution a Gaussian profile that measures thermal line broadening and a Lorentzian profile that measures nonthermal broadening. 
Therefore, a broad absorption feature may be due to high gas temperatures, pressures, or velocity dispersions. We note that although cosmic rays are a source of nonthermal pressure support they do not explicitly affect the Doppler-$b$ parameter value. Instead, they indirectly influence the Doppler-$b$ parameter by altering the density, temperature, velocity, or spatial extent of the gas.

The left panel of \autoref{fig:ovi_bval} shows the \ovi\ column density as a function of Doppler-$b$ parameter for \po\ and \pocr. Each scattered point represents a single \ovi\ absorption feature and is colored by whether or not that absorber is kinematically aligned with a \si\ feature (has a matched \si\ absorber within 35 km/s in line-of-sight velocity space). \po\ has significantly more ``no-lows" than either \pocr\ or the observations. This is in part due to the fact that \po\ has both more sightlines that have no detected \si\ and more sightlines with detected \si\ that is kinematically unrelated to \ovi. 

In the cosmic-ray-pressure-supported CGM of \pocr, the distribution of \ovi\ Doppler-$b$ values is narrower, likely because \ovi-bearing gas in \pocr\ tends to have cooler temperatures than \ovi-bearing gas in \po\ (\autoref{fig:phase}). \pocr\ also reproduces the observed abundance of \ovi\ Doppler-$b$ parameters that are $20-30$ km/s and predicts a relative gap in \ovi\ Doppler-$b$ parameters around 40 km/s. In comparison, \po\ has broader \ovi\ features, with the most abundant \ovi\ Doppler-$b$ parameter in the $30-40$ km/s range. 

Neither simulated distribution reproduces the very broad \ovi\ features seen in \cite{Werk:2016}\footnote{There is one additional \ovi\ absorber with $b = 174.7$ km/s in the \cite{Werk:2016} data that isn't pictured in \autoref{fig:ovi_bval}.}. The lack of very broad \ovi\ absorbers is likely influenced by the lack of \ovi\ in the outskirts of the simulated CGM and the simulation resolution, which narrows the velocity distribution within an \ovi\ cloud. The broad \ovi\ absorbers may also be affected by the small impact parameters of our sample or potential discrepancies in modeling the COS line-spread function in synthetic spectra.

\autoref{fig:siiii_bval} repeats the format of \autoref{fig:ovi_bval} for \si\ absorbers. In the left panel, each scatter point represents individual \si\ absorbers, colored by whether they have a corresponding \ovi\ absorber within 35 km/s in line-of-sight velocity space. The CGM of \pocr\ produces higher column densities of \si, which appear consistent with the range of observed column densities. However, it is difficult to compare precisely, as many of the observed \si\ absorbers are saturated. \pocr\ also produces distinctly higher \si\ Doppler-$b$ parameters that better agree with the observed distribution. However, as with \ovi\ absorbers, neither simulation reproduces the broad tail of \si\ Doppler-$b$ parameters.

We use a two-sample Kolmogorov-Smirnov (KS) test to investigate whether either \po\ or \pocr\ produces ion Doppler-$b$ parameters that sample the same distribution as the COS-Halos observations. In all cases, the KS statistic is inconclusive, and neither simulation produces an obviously better fit to the data. Besides differences in physical models, there are a variety of reasons why the simulations might not reproduce the observed Doppler-$b$ distributions, including resolution, the fact that the synthetic spectra only probe the inner CGM, or comparing the properties of a single galaxy, at a single point in time, against an observed variety of galaxies.

\subsection{The Kinematics of Cool and Warm Gas}
Next, we consider the impact of cosmic rays on the kinematics of warm and cool CGM gas traced by \ovi\ and \si. Using synthetic spectra, we compare the velocity offsets of individual absorbers of \ovi\ and \si\ and compare them directly against the properties of the velocity offsets of the absorbers in \citet{Werk:2016}.

In \autoref{fig:ovi_dvel}, we plot the histograms of minimum difference in velocity centroids between \ovi\ and \si\ for all sightlines that have detections of both ions. \po\ has significantly more spectra that have \si\ absorbers that are kinematically offset from the \ovi\ absorbers. This is because the cool, \si\ absorbing, gas in \po\ is often physically unrelated to the warm, \ovi\ absorbing, gas. By comparison, \pocr\ has a high incidence of kinematic alignment between \ovi\ and \si. 

As in the previous section, we use two-sample KS tests to see whether the simulated data is drawn from the same sample as the observed data. However, we again find that the KS tests are inconclusive.

The increased kinematic alignment between gas phases in \pocr\ comes from the overall effect of cosmic-ray pressure on the CGM kinematics.
\autoref{fig:phase_temperature} highlights the radial velocity differences between cool and warm gas in the inner CGM of both simulations. In \po, the majority of cool gas is radially infalling. Even if we were to exclude the fast inflowing cool gas that originates from the accreting stream, the remaining cool gas radial velocity histogram is still offset from the warm gas radial velocity histogram. Meanwhile, the cosmic-ray pressure support in \pocr\ helps support cool gas against gravity, giving it a very similar radial velocity distribution as that of the warm gas. 
In this way, significant cosmic-ray pressure support naturally leads to kinematic alignment between the cool and warm gas phases.

\subsection{Relating Kinematic Alignment to Physical Absorbers}
With \textit{synthetic} spectra, we have the unique opportunity to relate the kinematic alignment between \ovi\ and \si\ to the spatial relationship of the absorbing gas along the line of sight. To do this, we turn back to the simulated properties of each 1D sightline from which the synthetic spectra are generated. 

In our spectra, we see three physical explanations of kinematic alignment. (1) Some kinematic alignment is happenstance and comes from physically unrelated gas clouds. This physically unrelated kinematic alignment accounts for 37.5\% of \ovi-\si\ absorbers in \po\ and only 5\% of \ovi-\si\ aligned absorbers in \pocr. 

(2) Some kinematic alignment traces multiphase gas clouds -- either a cool cloud embedded in a warm cloud, or an intermediate temperature cloud that produces sufficient cool and warm components. These types of clouds tend to produce small kinematic offsets and comprise 62.5\% of aligned \ovi-\si\ absorbers in \po\ and 38\% percent of aligned \ovi-\si\ absorbers in \pocr. It is worth mentioning that we also see instances of \ovi\ and \si\ absorbers that come from the same physical cloud but that are kinematically misaligned. 

(3) The last physical explanation for kinematic alignment is when both cool and warm gas are tracing the large scale rotation of the CGM (for example, see the right panel of \autoref{fig:spectra_cr}). In this case, the warm gas envelops the inner cool gas, and this type of physical alignment produces slightly larger velocity offsets than clouds that are both kinematically and physically aligned. Roughly 67\% of kinematic alignment in \pocr\ falls into this category but is not present in \po.

In \autoref{fig:wass}, we quantify the offsets between line-of-sight velocities and positions of \ovi- and \si-bearing gas using the Wasserstein distance (also known as the earth mover's distance). The Wasserstein distance,  $W_p(m, n)$, measures the amount of ``work" it takes to transform one 1D distribution ($m$) into another ($n$),

\begin{equation}
    W_p(m, n) = \int_0^1 |F^{-1}_m(q) - F^{-1}_n(q)|^p\, {\rm d}q,
\end{equation}
where $F_m(x),\, F_n(x)$ are the cumulative distribution functions of $m$ and $n$. We assume $p = 1$. 

When defining the line-of-sight velocity and position samples, we consider all particles whose \ovi/\si\ column density is at least $10^{-2}$ of the total \ovi/\si\ column density of the sightline. We then weight each sample using the particles' fractional column density.

Within a single sightline, if the \ovi-bearing gas tends to have similar velocities as \si-bearing gas, then the Wasserstein distance between the two velocity distributions, $W_p(\rm{v}_{\rm OVI}, \rm{v}_{\rm SiIII})$, will be small. Similarly, if \ovi-bearing gas tends to be co-spatial with \si-bearing gas, then the Wasserstein distance between the two line-of-sight positions, $W_p(\rm{x}_{\rm OVI}, \rm{x}_{\rm SiIII}$, will be small. Conversely, if cool and warm gas is physically or kinematically unrelated, then the relevant Wasserstein distance(s) will be large. Considering an ensemble of sightlines, a tightly-correlated relationship between the Wasserstein distance for the velocities and spatial offsets would suggest that kinematic alignment in spectra also implies that \si- and \ovi-absorbing gas is co-spatial. Conversely, a complete lack of correlation would imply that kinematic alignment is not related to the physical location of the absorbing gas.

The scattered points in \autoref{fig:wass} show the Wasserstein distances of the line-of-sight velocities and positions of \ovi- and \si-bearing gas for all sightlines that contain detectable absorption features from both \ovi\ and \si. The color of each scattered point represents the impact parameter of that sightline.

In \po, there is no clear relationship between the kinematic and spatial alignment of cool and warm gas, traced by \ovi\ and \si. This result suggests that kinematic alignment of multiphase gas in \po\ spectra is unlikely to imply a physical relationship between the gas phases. We note that the lack of scattered points with high impact parameters is due to the relative lack of \si\ absorbers at large distances from the galaxy center. 

In the cosmic-ray-pressure-supported halo of \pocr, there is a positive correlation between the kinematic and spatial alignment of cool and warm gas. In particular, sightlines with impact parameters between 35-50 kpc have the smallest kinematic and spatial offsets, implying that \ovi\ and \si\ are arising from the same physical structures. Meanwhile, sightlines that directly pierce the central region of the \si\ structure have larger spatial and velocity offsets with \ovi-bearing gas. This is likely because the inner CGM \pocr\ is filled with a continuous \si-absorbing cloud that is pierced by smaller \ovi\ absorbers. The data from \pocr\ predict that small kinematic offsets ($< 10$ km/s) between \ovi\ and \si\ absorbers imply small spatial offsets ($< 10$ kpc). 

\section{Discussion}\label{sec:discussion}
\subsection{Implications for CGM Kinematics}
In this work, we demonstrate that the dominant source of pressure support (thermal or cosmic-ray) in the CGM alters the relationship between cool and warm gas kinematics.

We first consider the case of a thermal-pressure-supported CGM, like the one present in \po. In order to maintain thermal pressure equilibrium, cool gas must have significantly higher densities than warm gas.
Therefore, cool gas forms small, dense clouds that are not buoyant in the CGM. The lack of buoyancy implies that regardless of its origin, cool gas will accelerate towards the galactic disk. Even if cool gas initially shares the velocity of the warm phase (for example, cool gas that forms through thermal instability), the common velocity phase will be short-lived as cool droplets will accelerate towards the galaxy with a characteristic timescale set by the gravitational 
freefall time, $t_{\rm ff} = 2 r / v_{\rm c}$. For the inner CGM ($r = 50$ kpc) of a MW-sized galaxy, with a circular velocity of $v_{\rm c} = 200$ km/s, the freefall time is roughly $t_{\rm ff} = 500$ Myr. Therefore, in a thermal-pressure-supported CGM, one would expect a large fraction of cool gas absorbers to have at least some velocity offsets from warm absorbers. 

In a cosmic-ray-pressure-supported CGM, the cool gas phase has significantly lower densities and is supported against gravity by the cosmic-ray pressure gradient. Rather than contracting into into dense clouds, low-density cool gas fills the entire inner CGM, out to $\sim50$ kpc. The combination of a cosmic-ray pressure gradient contributing to maintaining hydrostatic equilibrium and the increased buoyancy of the low-density cool gas prevents cool gas from sinking as rapidly in the gravitational potential well \citep{Butsky:2020}. Together, these effects result in cool gas having a similar velocity distribution to the warm gas phase. This naturally leads to a kinematic alignment between multiphase ions in absorption spectra. 

In both galaxy models, the radial velocity distributions of warm gas are similar to the radial velocities of hot gas (\autoref{fig:phase_temperature}). Therefore, we would expect absorption from higher ionization states of oxygen (O~{\sc vii}, O~{\sc viii}) detected in future X-ray absorption studies to be kinematically aligned with \ovi\ absorbers. 

\subsection{The Relationship Between Kinematic and Spatial Alignment}
In addition to altering CGM kinematics, the dominant source of pressure support (thermal or cosmic-ray) in the CGM alters the relationship between kinematic alignment and physical gas structure. 

In the thermal-pressure-supported CGM of \po, gas positions and velocities are generally uncorrelated. Although in some cases, kinematic alignment does correspond with spatial alignment, we did not discover any trends that could discern such cases from coincidental kinematic alignment. 
In the cosmic-ray-pressure-supported CGM of \pocr, kinematic alignment \textit{does} correlate with spatial alignment. Absorbers with small velocity differences tend to come from the same gas structures. Spectra in which \ovi\ absorption features were nestled inside a broad \si\ absorption feature traced the rotation of the extended disk in which \ovi\ gas enveloped \si\ gas. 

In both galaxy models, the kinematics of cool gas is particularly decoupled from
gas positions in the case of a coherent inflowing stream. Therefore, the more we expect coherent inflows to be an important component of CGM structure, the less we should trust velocity alignment as an indicator of spatial alignment. 

\subsection{Uncertainty in Modeling Cosmic Rays}
It is now well-established that cosmic rays are produced in supernova shocks \citep[e.g.,][]{Ackermann:2013} and are an important source of energy in the galactic disk \citep[e.g.,][]{Boulares:1990}. 
However, despite significant recent theoretical and computational advances in cosmic-ray hydrodynamics \citep[e.g.,][]{Jiang:2018, Thomas:2019, Hopkins:2021_crm1}, there is no consensus as to which existing cosmic-ray transport models (if any) realistically model the true behavior of cosmic rays.
Several recent studies have demonstrated that different cosmic-ray transport models can change the predicted properties of the galactic disk, galactic outflows, and CGM structure \citep[e.g.,][]{Ruszkowski:2017, Butsky:2018, Buck:2020, Hopkins:2021_transport1}. Additionally, even the same cosmic-ray transport physics can produce significant differences in galaxy properties due to differences in numerical implementation \citep[e.g.,][]{Gupta:2021, Semenov:2021}.

In this work, we model cosmic-ray transport as isotropic diffusion, which is an oversimplification of the true microphysical processes that govern cosmic-ray transport. There are several limitations to such an approach.  For example, in reality, cosmic-ray transport is confined to motion around magnetic field lines and is inherently anisotropic \citep{Cerri:2017, Evoli:2018}. Modeling cosmic-ray transport as isotropic can affect the quantitative details of the simulated galaxy, for example, by overestimating the strength of galactic outflows \citep[e.g.,][]{Pakmor:2016}. 

Additionally, we neglect cosmic-ray streaming, which may be an important form of cosmic-ray transport \citep{Evoli:2018, Thomas:2020}. A key difference of the streaming approximation is an Alfv{\'e}n wave cooling term that transfers cosmic-ray energy into heating the thermal gas. By omitting this term, the diffusion approximation likely overestimates the amount of cosmic-ray energy that escapes the galactic disk \citep[e.g.,][]{Wiener:2017, Buck:2020}. However, this heating term is weak relative to gas cooling in the CGM \citep{Ji:2020}. In the streaming approximation, a build up of cosmic-ray pressure at cold cloud boundaries may accelerate cold gas clouds via the ``bottleneck'' effect \citep{Skilling:1971, Wiener:2017_coldcloud, Wiener:2019, Thomas:2021}. While the ``bottleneck'' effect has the potential to alter the kinematic signature of cold CGM gas, it may be less effective in the context of a cosmic-ray-pressure-supported halo that is filled with low-density cool gas as opposed to individual dense cloudlets. 

Finally, in our simulations, we model cosmic rays as a single fluid of GeV protons. Although this is a common assumption, explicitly simulating a wide range of cosmic ray energies would likely change the quantitative details of the cosmic-ray pressure support in the CGM \citep[e.g.,][]{Ensslin:2007, Jubelgas:2008, Hopkins:2022_crspectra, Girichidis:2022}

The field of cosmic-ray hydrodynamics is rapidly evolving, and existing state-of-the-art cosmic-ray physics implementations that can simultaneously model both diffusion and streaming \citep[e.g.][]{Jiang:2018, Chan:2019, Thomas:2019, Girichidis:2022} may be misrepresenting the true behavior of cosmic rays in the CGM. For example, recent studies demonstrated that neither streaming nor diffusion can reproduce the the breadth of observed cosmic-ray scaling relations in the Solar circle \citep{Hopkins:2022,Kempski:2022}, hinting at a significant gap in our understanding of the microphysical processes that drive cosmic-ray transport on galactic scales.

In the face of these extreme uncertainties in modeling cosmic-ray transport, we expect many of the quantitative details of our results, for example the exact ion column densities, the radial extent of the cosmic-ray-pressure-supported region, or the details of the gas phase distribution to change with different cosmic-ray transport models. However, we expect the qualitative details, in particular the relatively cool \ovi\ absorption, the low-density \si\ gas, and the increased kinematic alignment between gas phases to be robust for a CGM that is primarily supported by nonthermal pressure. While a variety of cosmic-ray transport models can produce a cosmic-ray-pressure-supported CGM \citep[e.g.][]{Salem:2016, Butsky:2018, Buck:2020, Ji:2020} that is consistent with $\gamma$-ray observational constraints \citep{Chan:2019}, those same models are also consistent with a wide range of more modest CGM cosmic-ray pressure profiles. In the case of a CGM with cosmic-ray pressures equal to or less than the thermal pressure, we expect qualitative properties that are ``in-between'' those of a thermally-pressure-supported CGM and a cosmic-ray-pressure-supported CGM.

\subsection{Uncertainty in Cold Gas Structure}
One of the largest existing uncertainties about CGM structure is the physical state of cold gas. Recent theoretical and numerical works suggest that the characteristic scale of cold gas (in the absence of significant nonthermal pressure support) is on the order of parsecs, depending on the local gas sound speed and cooling time \citep{McCourt:2018, Li:2020}. However, whether cold gas exists as a mist of small cloudlets or whether those cloudlets coagulate to form larger clouds depends on the local gas properties \citep[e.g.][]{ Gronke:2018, Gronke:2020, Sparre:2020, Kanjilal:2020, Abruzzo:2021, Farber:2022}. It is not yet computationally feasible to directly resolve parsec scales in throughout the CGM of simulated galaxies, and studies that push the boundary of resolution capabilities find that cold gas properties are not converged in galaxy-scale simulations \citep[e.g.,][]{Hummels:2019, Peeples:2019, Suresh:2019, vandeVoort:2019, Mandelker:2021}.

If cold gas forms a mist in a thermal-pressure-supported CGM, then turbulent motions may efficiently counteract gravity and reduce the radial velocity differences between the cold and hot phases. However, if cold gas is likely to coagulate into larger clouds, then we expect a more pronounced difference in the radial velocity distributions of cold and hot gas. Additionally, the kinematics of cold gas may be strongly influenced by momentum transfer from the hot gas through its turbulent boundary layer \citep{Fielding:2020}. This boundary layer would contain intermediate-temperature gas with its own turbulent velocity structure that may alter the absorption signatures and kinematic alignment of cold gas in synthetic spectra.

In the absence of sufficient resolution to directly model the characteristic scale of cold gas clouds and the turbulent mixing layers at their boundaries, galaxy-scale simulations are likely to produce resolution-limited dense cold clouds that are not buoyant in a thermal-pressure-supported CGM. As we discuss more in \autoref{sec:appendix_GM}, we expect the relative radial velocity offset between cold and hot gas to be a persistent feature of simulated low-redshift CGM around L$^*$ galaxies. However, determining whether this offset is an inherent property of cold CGM gas in a thermal-pressure-supported CGM will first require a deeper understanding of the small-scale physics that governs cold gas evolution.

\section{Summary} \label{sec:summary}
In this work, we use synthetic spectroscopy to investigate the role of cosmic-ray pressure in setting the observed absorption-line properties and the kinematics of cool and warm gas in the CGM of low-redshift L$*$ galaxies. We simulate two Milky Way-sized galaxies, one without cosmic rays (\po), and one with cosmic-ray supernova feedback (\pocr). By a redshift of $z = 0.25$, these two galaxies, which started from the same initial conditions, have evolved fundamentally different CGM properties. 

\po\ has a hot, thermal-pressure-supported CGM that is pierced by cool cloudlets in the inner CGM as well as an accreting stream from a recent merger event. By contrast, the inner CGM of \pocr\ is supported against gravity by cosmic-ray pressure and filled with cool, low density gas. 

We systematically probe the inner CGM of both galaxies with synthetic spectra and compare the extracted column densities, Doppler-$b$ parameters, and velocity centroids directly against COS-Halos data. We focus our analysis on two ions: \ovi, which traces warm $\sim 10^{5.5}$ K gas, and \si, which traces cool $\sim 10^{4.2}$ K gas. 

Our results are summarized as follows. 

(1) A cosmic-ray-pressure-supported CGM results in more narrow \ovi\ Doppler-$b$ parameters than a thermal-pressure-supported CGM. This is due to the relatively cooler temperatures of \ovi-bearing gas in the \pocr\ simulation. In \pocr, \ovi\ absorption peaks at $10^{5.5}$ K, while in \po, \ovi\ traces both warm and hot gas with temperatures of $10^{5.5-5.9}$ K (\autoref{fig:phase}). While \pocr\ can explain the observed relative abundance of \ovi\ Doppler-$b$ parameters between 20-30 km/s, neither simulation offers a statistically good fit to the observed data. Notably, neither simulation reproduced the observed large \ovi\ Doppler-$b$ parameters $> 60$ km/s (\autoref{fig:ovi_bval}). 
   
{(2) A cosmic-ray-pressure-supported CGM produces broader \si\ Doppler-$b$ parameters.} In both simulations, \si\ traces gas with temperatures around $10^{4.2}$ K (\autoref{fig:phase}). In \po, cool gas tends to form small cloudlets with narrow Doppler$-b$ parameters (\autoref{fig:siiii_bval}). In contrast, the low-density cool gas in \pocr\ fills the entire inner CGM, resulting in large column densities and broader absorption features of \si. While the distribution of \si\ Doppler-$b$ parameters in \pocr\ is in better agreement with observations, it still underpredicts the observed abundance broad \si\ features. 
    
{(3) A cosmic-ray-pressure-supported CGM predicts substantially more kinematic alignment of low and high ions.} There is significantly more kinematic alignment between \ovi\ and \si\ in \pocr\ (\autoref{fig:ovi_dvel}). We demonstrate that in a cosmic-ray-pressure-supported halo, in which low-entropy cool gas is supported against gravity, the velocity distribution of cool and warm gas is similar, leading to a natural alignment between multiphase ions. This contrasts the kinematics of the thermally supported CGM of \po, in which cool gas is denser and is more likely to have an inward radial velocity than the warm gas (\autoref{fig:phase_temperature}). 

{(4) A cosmic-ray-pressure-supported CGM predicts a relationship between kinematic and physical alignment.} We demonstrate that kinematic alignment does not necessarily imply that the absorbers trace physically related clouds. Our simulations predict that in a cosmic-ray-pressure-supported CGM, there is a correlation between the kinematic offset between \ovi\ and \si\ absorbers and their physical separation along the line of sight (\autoref{fig:wass}). By contrast, kinematic and spatial alignment in \po\ are uncorrelated. 

While the quantitative details of modeling cosmic rays in galaxy-scale simulations are still poorly-constrained, our study highlights how significant cosmic-ray pressure support in the CGM fundamentally changes the kinematic signatures of multiphase gas. Ultimately, this work demonstrates that detailed comparisons between simulations and observations using synthetic spectroscopy are a powerful tool for constraining the dominant physical processes that govern galaxy evolution.

\acknowledgements
The authors thank Philip Hopkins, Suoqing Ji, and Peng Oh, and the anonymous referee for their insightful comments. IB was supported by HST theory grant HST-AR-15046, the DuBridge Postdoctoral Fellowship from Caltech, and by her tenure as a Blue Waters Graduate Fellow. The Blue Waters sustained-petascale computing project is supported by the National Science Foundation (Grants No. OCI-0725070 and No. ACI-1238993) and the State of Illinois. JKW acknowledges support as a Cottrell Scholar, from the Research Corporation for Science Advancement, grant ID number 26842. Additionally, JKW and KT acknowledge support for this work from NSF-AST 1812521. DBF, through the Flatiron Institute, is supported by the Simons Foundation. This research benefited from the KITP Program: ``Fundamentals of Gaseous Halos'', and thus was supported in part by the National Science Foundation under Grant No. NSF PHY-1748958.

\bibliography{main}

\appendix
\label{sec:appendix}
\begin{figure*}
\includegraphics[width=\textwidth]{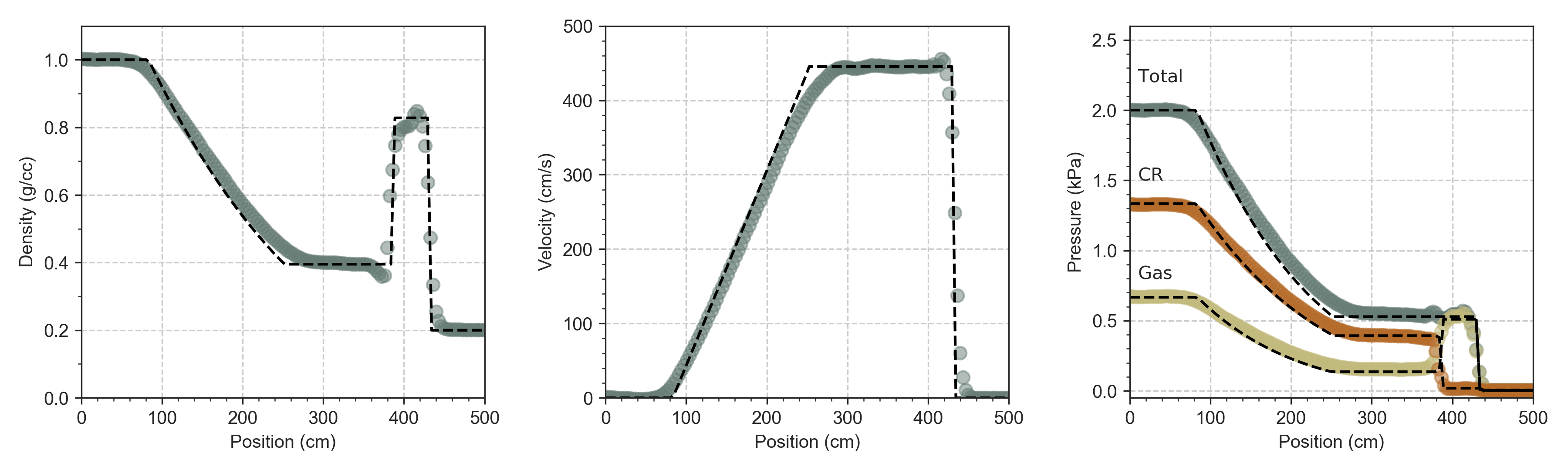}
\caption{ The SOD shocktube. The state of the gas density (left), velocity (middle), and pressure (right) at $t = 0.31$ seconds. The dashed line shows the analytic solution for the gas density, velocity, and total pressure, which is a sum of gas and cosmic-ray pressures. The shocktube was intialized with a contact discontinuity at $x = 250$ cm at $t = 0$. To the left of the discontinuity, the gas density, pressure, and cosmic-ray pressure are initialized to $\rho = 1 {\rm g/cm}^3$, $P_{\rm g} = 2/3$ kPa, $P_{\rm c} = 4/3$ kPa. To the right of the discontinuity, the gas density, pressure, and cosmic-ray pressure are initialized to $\rho = 0.2 {\rm g/cm}^3$, $P_{\rm g} = 267.2$ Pa, $P_{\rm c} = 267.2$ Pa. The velocity is initialized to zero everywhere. }
\label{fig:sod} 
\end{figure*}

\section{Cosmic Rays in ChaNGa}\label{sec:appendix_changa}
\subsection{Equations of Motion}
In the Lagrangian approximation, a fluid is discretized by some number of massive particles. Traditionally, each particle has a mass, velocity, and specific energy, and these quantities are evolved from their initial conditions following conservation equations. We model cosmic rays as an additional ``fluid" such that each particle also contains a specific cosmic-ray energy. The evolution of these particles is governed by the equations for the conservation of mass, momentum, thermal energy, and cosmic-ray energy below.

\begin{equation}
\frac{D\rho}{Dt} = -\rho\nabla\cdot{\bf v}
\end{equation}

\begin{equation}
\frac{\rho D  {\bf v}}{D t} + \nabla(P_{\rm tot}) = {\bf g}
\end{equation}

\begin{equation}
\frac{\rho D u_{\rm g}}{D t} + P_{\rm g} \nabla\cdot{\bf v} = \Lambda_{\rm g} - \Gamma_{\rm g}
\end{equation}

\begin{equation}
    \frac{\rho D u_{\rm c}}{D t} + P_{\rm c} \nabla\cdot{\bf v} = \nabla \cdot(\kappa_{\rm c}\nabla u_{\rm c}) + \Lambda_{\rm c}.
\end{equation}
We define $\rho, {\bf v}, u_{\rm g}$ to be the gas density, velocity, and specific internal energy (energy per mass). Similarly, $u_{\rm c}$ is the cosmic-ray specific internal energy. The gas and cosmic-ray pressures are related to the internal energy by the adiabatic index, $P_{\rm g} = (\gamma - 1) \rho u_{\rm g}$ and $P_{\rm c} = (\gamma_{\rm c} - 1) \rho u_{\rm c}$, where $\gamma = 5/3$ and $\gamma_{\rm c} = 4/3$. The total pressure of the system given by the sum of thermal and cosmic-ray pressures, $P_{\rm tot} = P_{\rm g} + P_{\rm c}$. ${\bf g}$ is the gravitational force. $\Lambda$ and $\Gamma$ are shorthand representations of the total source and sink terms of gas and CR energy. $\kappa_{\rm c}$ is the constant cosmic-ray diffusion coefficient. Lastly, $t$ represents time, and $D/Dt = \partial/\partial t + {\bf v} \cdot \nabla$ denotes the convective derivative. 

%Practically, these equations are implemented in the code as the following algorithms. 
In the SPH approach, the above equations become ordinary differential equations for the motion, internal energy, and cosmic-ray energy of each particle as follows:

\begin{equation}
\frac{d{\bf v}_{\rm i}}{d t} = - \sum_{\rm j} m_{\rm j} \bigg(\frac{P_{\rm tot,i} + P_{\rm tot, j}}{\rho_{\rm i} \rho_{\rm j}} + \Pi_{\rm ij}
\bigg) \nabla_{\rm i} \bar{W}_{\rm ij}
\end{equation}

\begin{equation}
\frac{d u_{\rm g, i}}{d t} = - \sum_{\rm j} m_{\rm j} \bigg(\frac{P_{\rm g,i} + P_{\rm g,j}}{\rho_{\rm i} \rho_{\rm j}} + \Pi_{\rm ij}
\bigg) \nabla_{\rm i} \bar{W}_{\rm ij}
\end{equation}

\begin{equation}
\frac{d u_{\rm c, i}}{d t} = - \sum_{\rm j} m_{\rm j} \bigg(\frac{P_{\rm c,i} + P_{\rm c,j}}{\rho_{\rm i} \rho_{\rm j}} + \Pi_{\rm ij}
\bigg) \nabla_{\rm i} \bar{W}_{\rm ij}.
\end{equation}
In the equations above, $m$ is the mass of an individual particle, $\Pi$ is an artificial viscosity term and $W$ is the general SPH kernel function. The subscript $i$ indicates a single particle, and $\sum_{\rm j}$ is the sum of properties over that particle's nearest neighbors. Lastly, we model the diffusion of cosmic-ray energy as

\begin{equation}
    \frac{d u_{\rm c,i}}{d t}|_{\mathrm{Diff}} = -\sum_{\rm j} m_{\rm j} \frac{\kappa_{\rm c}(u_{\rm c,i} - u_{\rm c,j})({\bf r}_{\rm ij} \cdot \nabla_{\rm i} \bar{W}_{\rm ij})}{\frac{1}{2} (\rho_{\rm i} + \rho_{\rm j}){\bf r}^2_{\rm ij}},
\end{equation}
where ${\bf r}_{\rm ij}$ is the distance between neighboring particles. We solve diffusion using an explicit scheme with a timestep constraint, 
\begin{equation}
    \Delta t = 0.5 \eta_{\rm c} \frac{h^2}{\kappa_{\rm c}},
\end{equation}
where $\eta_{\rm c}$ is the constant Courant factor and $h$ is the SPH particle smoothing length. We ensure that the cosmic-ray (and thermal) energy is never negative by flagging particles that would develop negative energies at a given time step and instead having them exponentially decay to the predicted value as described by

\begin{equation}
    u_{\rm c, i+1} = u_{\rm c, i}e^{u_{\rm c, pred} / u_{\rm c, i}},
\end{equation}
where $u_{\rm c, pred}$ is the predicted negative value of the specific cosmic-ray energy.

We note that we do not explicitly include Coulomb and hadronic losses, therefore the cosmic-ray pressure in \pocr\ represents an upper limit to the cosmic-ray pressure support we would expect.

In the following section, we test the performance of advection and diffusion of the newly added cosmic-ray fluid against analytic solutions.

\section{Tests of CR Performance}
\label{sec:appendix_crtests}
\subsection{SOD Shock-tube}
We demonstrate the advection of the thermal and cosmic-ray fluids using a modified SOD shocktube test \citep{Sod:1978, Pfrommer:2006}. In the initial conditions, 24,000 particles are arranged in a long, narrow three-dimensional glass \citep{sphglass} with dimensions $100 \times 100 \times 1000$ cm and periodic boundary conditions. Smoothing is done with an M$_4$ cubic spline kernel using 64 neighboring particles. The initial discontinuity is placed at $z = 250$ cm. To the left of the discontinuity, the gas density, pressure, and cosmic-ray pressure are initialized to $\rho = 1 {\rm g/cm}^3$, $P_{\rm g} = 2/3$ kPa, $P_{\rm c} = 4/3$ kPa. To the right of the discontinuity, the gas density, pressure, and cosmic-ray pressure are initialized to $\rho = 0.2 {\rm g/cm}^3$, $P_{\rm g} = 267.2$ Pa, $P_{\rm c} = 267.2$ Pa. The velocity is initialized to zero everywhere. 

\autoref{fig:sod} shows the evolved state of the modified SOD shocktube after $t = 0.31$ seconds. The scattered points show the gas properties, averaged in spatial bins with $\delta x = 5$ cm. The black dashed line shows the analytic solution for the gas density, velocity, and total pressure. The simulated shocktube follows the analytic predictions well. The slight noise to the right of the contact discontinuity ($z \sim 400$ cm), is likely due to resolution-related particle noise.

\subsection{Diffusion}
\begin{figure}
\includegraphics[width=0.5\textwidth]{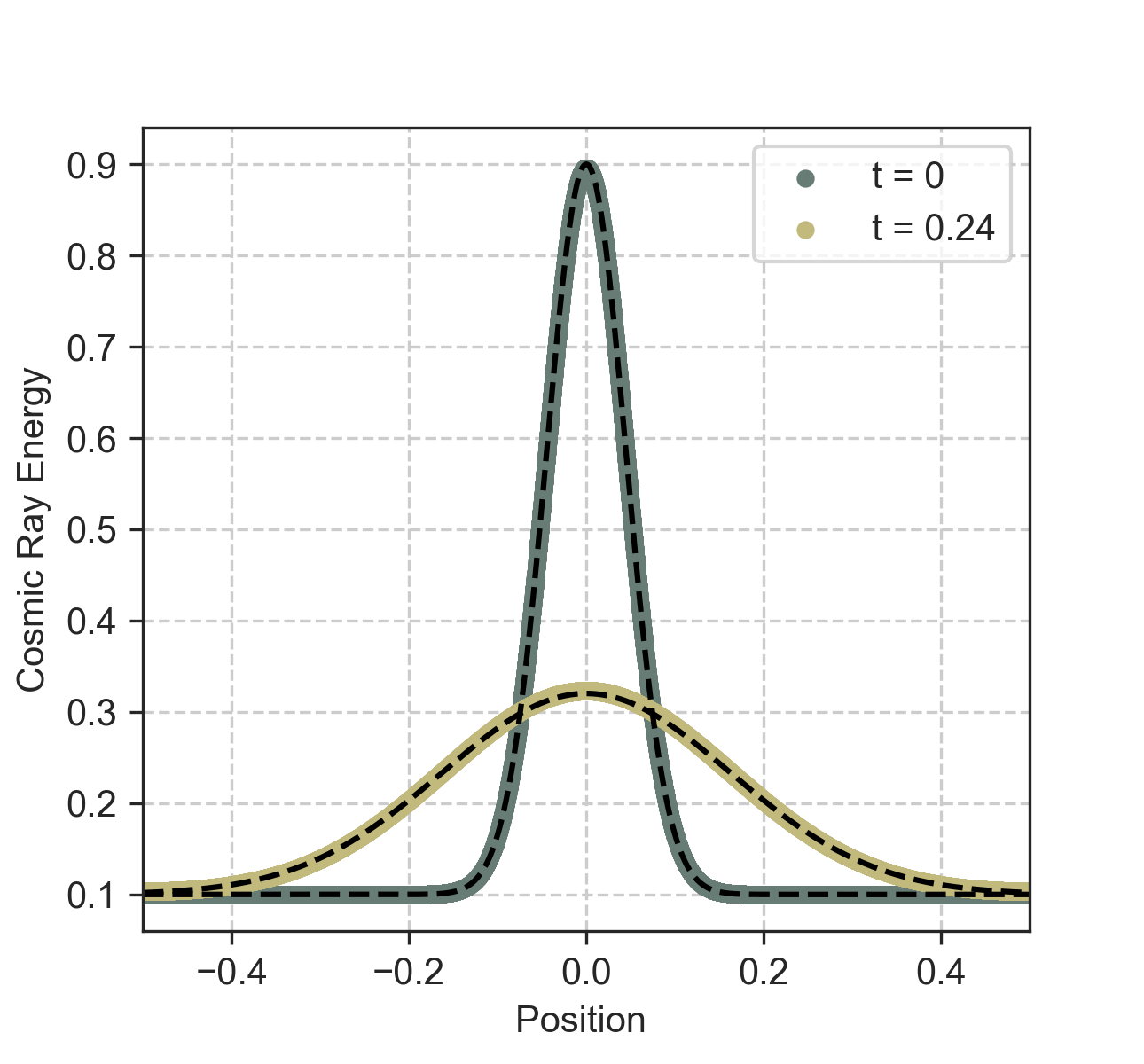}
\caption{ Demonstration of cosmic-ray diffusion. An initial Gaussian overdensity of cosmic-ray energy decays overtime as cosmic rays diffuse down their energy gradient with a constant diffusion coefficient. The y-axis shows radially-averaged values of the cosmic-ray energy as a function of position, $r = x^2 + y^2$. The black dashed lines show analytic solutions for $t = 0$ and $t = 0.24$ Myr. }
\label{fig:diffusion} 
\end{figure}

\begin{figure}
\includegraphics[width=0.5\textwidth]{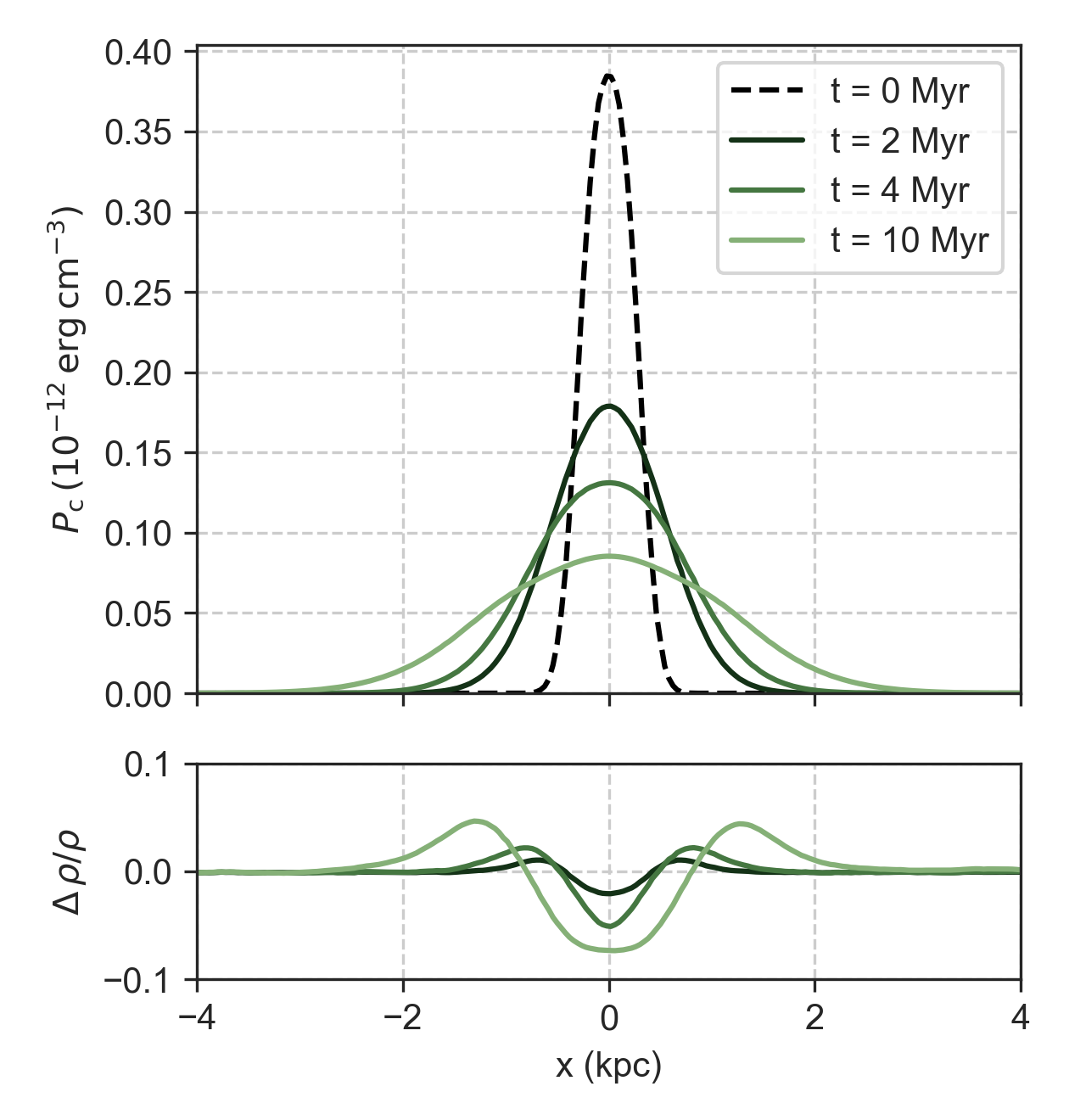}
\caption{The diffusion over time of an initial Gaussian overdensity of cosmic-ray pressure in a flux tube with gas motion turned on. We choose a diffusion coefficient of $\kappa_{\rm c} = 3\times 10^{28}\, {\rm cm}^2\, {\rm s}^{-1}$. The cosmic-ray pressure gradient generates small perturbations in the gas density that propagate outward at the sound speed. }
\label{fig:diffusion_withgas} 
\end{figure}

\begin{figure*}
\includegraphics[width=\textwidth]{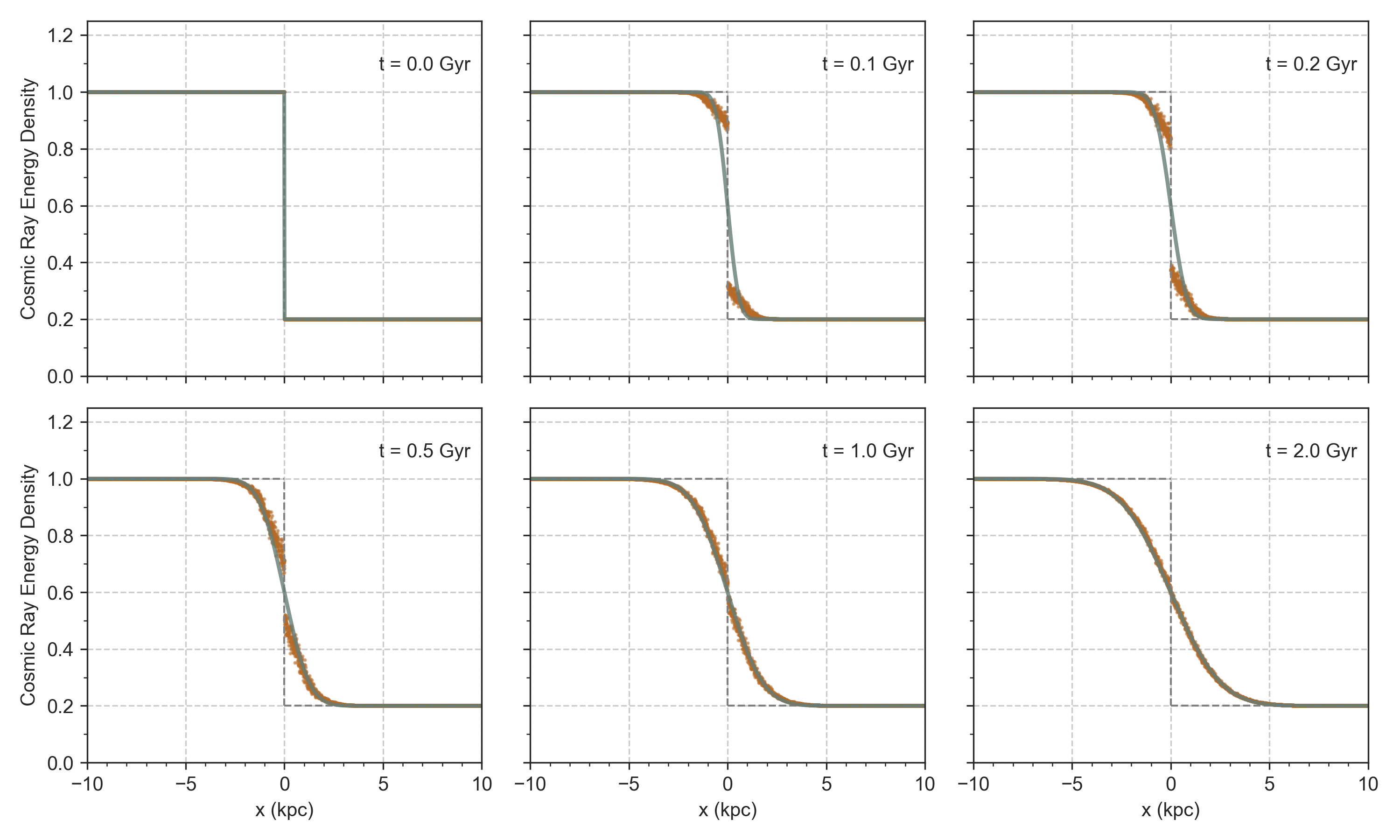}
\caption{The time evolution of cosmic-ray diffusion from an initial step function. The gray dashed line shows the initial conditions, the gray solid line shows the analytic solution given by \autoref{eq:jubelgas}, and the scattered orange points show individual simulated particles. We suppress the motion of gas particles to model only cosmic-ray diffusion with a constant diffusion coefficient $\kappa_{ c} = 1\, {\rm kpc}^2/{\rm Gyr}$. While at early times, the simulation deviates from the solution at the sharp discontinuity, it converges to the analytical solution at late times.}
\label{fig:jubelgas} 
\end{figure*}

In this work, we model cosmic-ray transport as isotropic diffusion with a constant diffusion coefficient, $\kappa_{\rm c}$. We test the implementation of cosmic-ray diffusion by measuring the evolution of an initial overdensity of cosmic-ray energy. We initialize 20,000 particles as an SPH glass with dimensions $0.2 \times 0.2 \times 1$ kpc. The distribution of cosmic-ray energy density is initialized as:
\begin{equation}\label{eq:gauss}
    u_{\rm c} = u_{\rm c,0}e^{-r^2 / 2D},
\end{equation}
where $r = x^2 + y^2$ is the distance from the center. We choose constants $u_{\rm c,0} = 0.9$, $D = 0.002$ and a diffusion coefficient of $\kappa_{\rm c} = 0.05$ kpc$^2/$Myr. In order to only test cosmic-ray diffusion in the absence of advection, we turn off gas particle motions.

\autoref{fig:diffusion} shows the evolution of an initial Gaussian overdensity of cosmic-ray energy assuming diffusion as the only form of cosmic-ray transport (neglecting any advection due to gas motion). Our implementation follows the expected analytic solution after $t = 0.24$ Myr. 

In \autoref{fig:diffusion_withgas} we demonstrate the behavior of cosmic-ray diffusion in a homogeneous flux tube with gas motions turned on. The background density in the flux tube is set to $\rho = 10^{26} {\rm g\, cm}^{-3}$ with an adiabatic sound speed of $c_s = 100\, {\rm km/s}$. The physical domain spans 4 kpc by 4 kpc by 10 kpc with periodic boundary conditions, resolved by $4 \times 10^4$ particles initialized in a glass configuration. The cosmic-ray energy is initially distributed as a Gaussian overdensity as described by \autoref{eq:gauss}, with $D = 0.2\, {\rm kpc}$. The normalization of the cosmic-ray profile is chosen so that at its maximum, the cosmic-ray pressure is equal to the gas pressure. We use a diffusion coefficient of $\kappa = 3\times 10^{28} {\rm cm}^2\, {\rm s}^{-1}$. 

The cosmic-ray pressure gradient initially accelerates small gas overdensities (as shown in the bottom panel) which then travel outward at the adiabatic sound speed. At later times, the cosmic-ray pressure profile maintains a nearly Gaussian distribution. The evolution of the cosmic ray pressure and gas density perturbations  are qualitatively similar to those found in a similar test in \citet{Wiener:2017}.

In \autoref{fig:jubelgas}, we demonstrate the behavior of cosmic-ray diffusion at a sharp discontinuity to gain insight into the behavior of cosmic-ray diffusion in phase transitions in the ISM and CGM. The initial conditions start with a three-dimensional glass with dimensions 2 kpc x 2 kpc x 20 kpc. The particles are initialized with uniform gas properties except for a sharp discontinuity in cosmic-ray energy at the center of the glass, where the cosmic-ray energy density jumps from 1.0 to 0.2 (in arbitrary units). We fix the motion of gas particles and demonstrate the diffusion of cosmic-ray energy at various snapshots over 2 Gyr. The dashed gray lines indicate the initial conditions while the solid gray lines indicate the time-dependent analytic solution as described in \citet{Jubelgas:2008}, 

\begin{equation}\label{eq:jubelgas}
    \varepsilon(x,t) = \begin{cases}
        \varepsilon_l + \frac{\varepsilon_r - \varepsilon_l}{2} \big [{\rm erf}\big (\frac{x}{\sqrt{4\kappa_{ c}t}}\big) + 1\big ] & {\rm for}\, x < 0 \\
        \varepsilon_r + \frac{\varepsilon_r - \varepsilon_l}{2} \big [{\rm erf}\big (\frac{x}{\sqrt{4\kappa_{ c}t}}\big) - 1\big ] & {\rm for}\, x \ge 0,
    \end{cases}
\end{equation}

where $\varepsilon$ is the cosmic-ray energy density, $\kappa_{ c}$ is the constant diffusion coefficient, $erf$ is the Gauss error function, t is the time, and the subscripts $l, r$ represent the initial conditions to the left and right of the discontinuity respectively.

At early times, the simulated values deviate from the analytic solution, characteristic of the behavior of SPH at sharp discontinuities. At late times, the simulation converges to the true solution. Given this behavior, our simulations likely mispredict the exact cosmic-ray energies at sharp boundaries, for example the injection of cosmic-ray energy by supernovae in the ISM or at the outskirts of the CGM as cosmic rays populate it for the first time. However, for the inner CGM at low-redshifts, we expect the cosmic-ray injection rate to be in a steady state and expect our numerical implementation to have converged to the analytic solution.

\section{Radial Velocity Profiles in Other Thermally-Supported Galaxies}\label{sec:appendix_GM}
\begin{figure}
\includegraphics[width=0.45\textwidth]{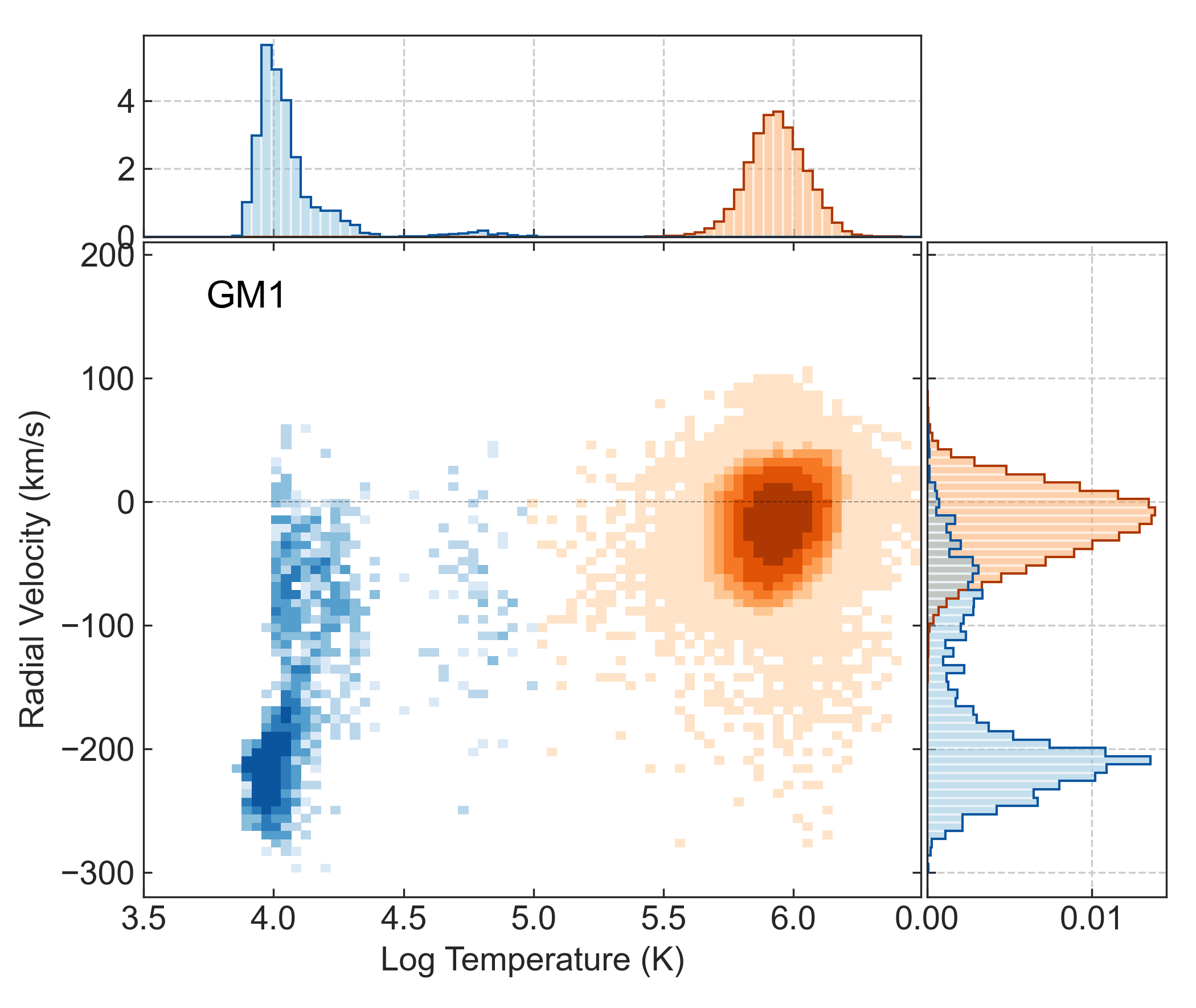}
\includegraphics[width=0.45\textwidth]{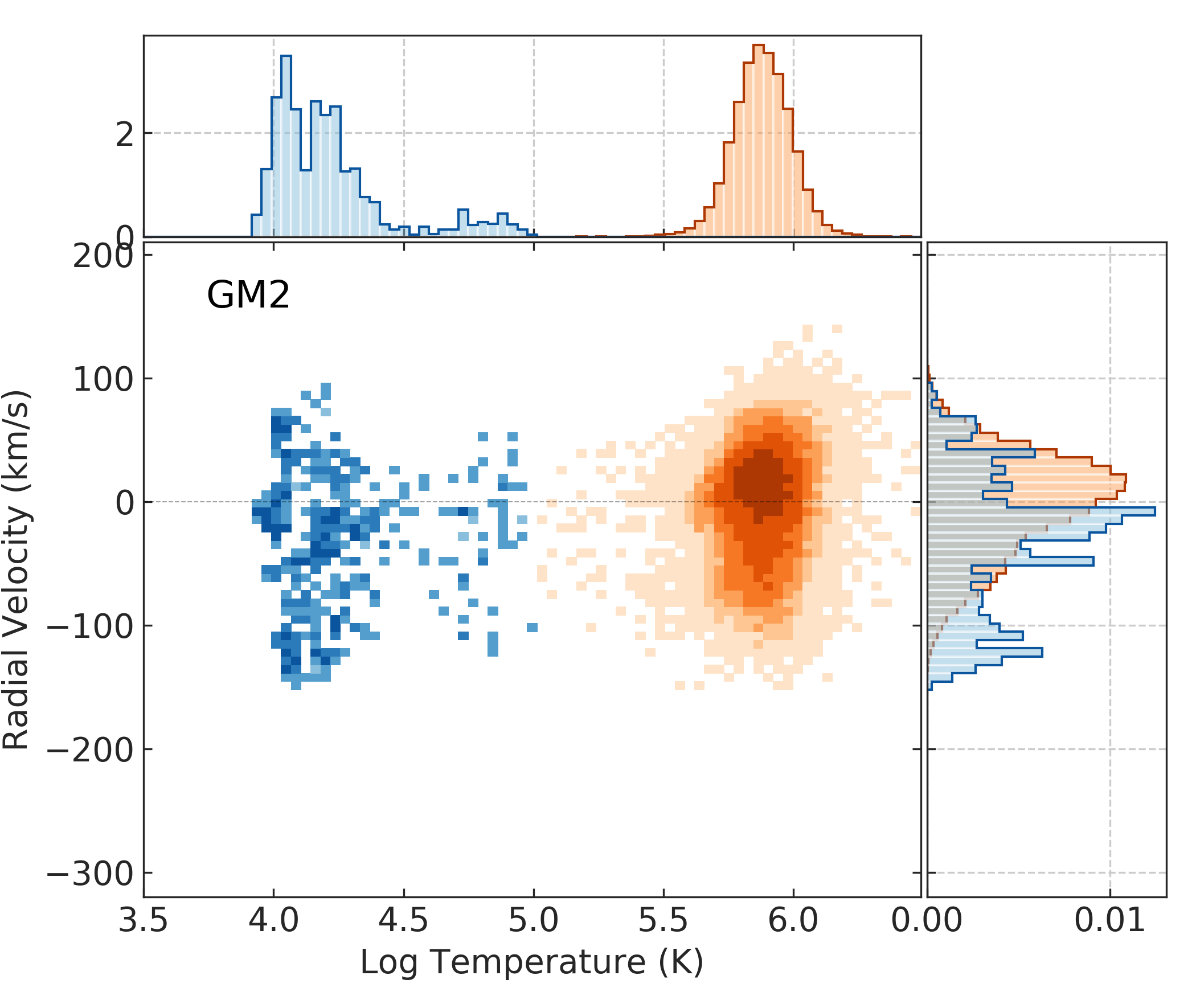}
\includegraphics[width=0.45\textwidth]{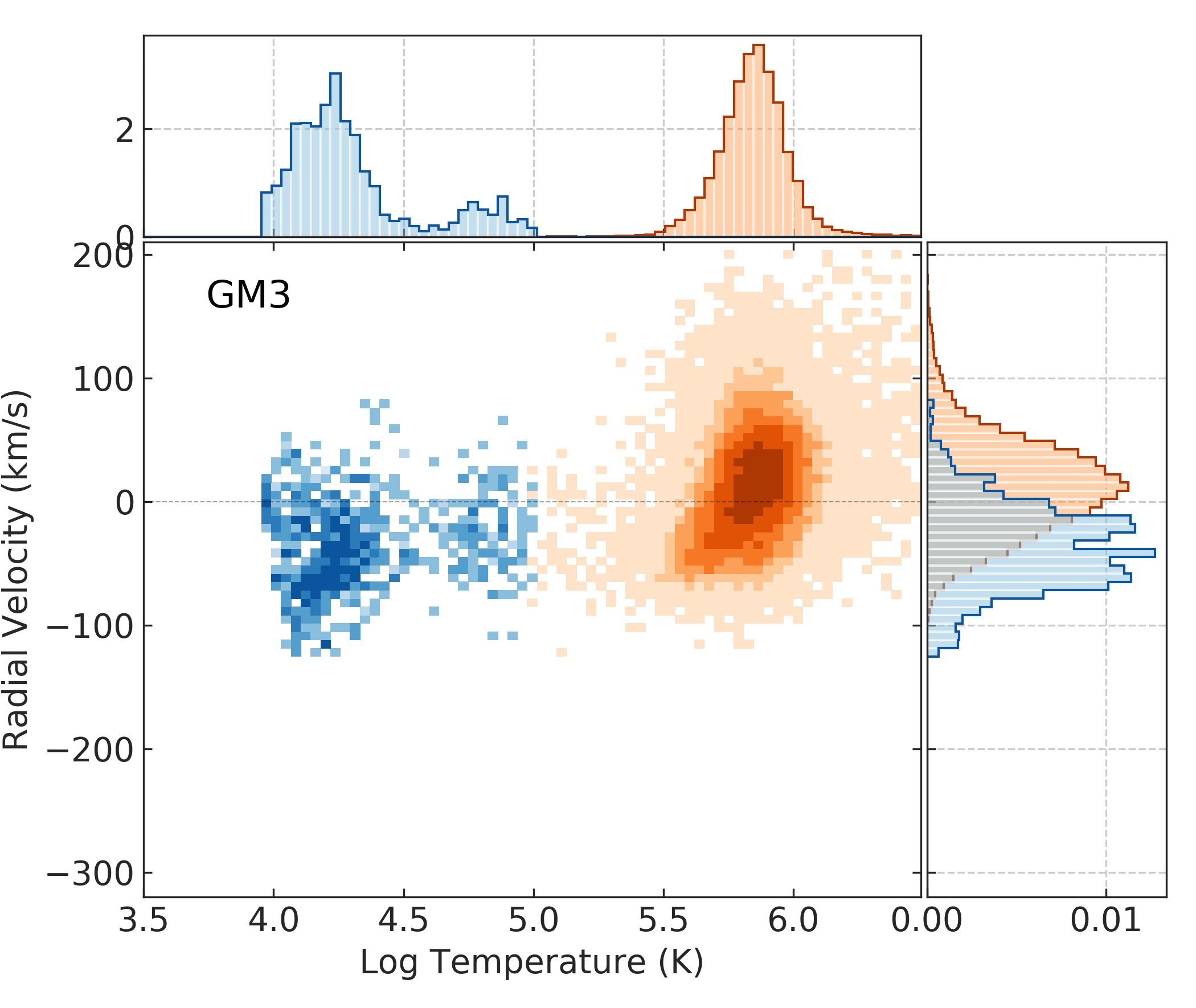}
\caption{The mass-weighted 2D histogram of radial velocity as a function of temperature for cool (blue) and warm (orange) CGM gas for three different galaxies, GM1, GM2, and GM3. The data is taken from gas that is between 10 and 50 kpc from the galactic center.} 
\label{fig:GM} 
\end{figure}

In \autoref{fig:GM} we explore the radial velocity -- temperature relationship in three additional galaxies, GM1, GM2, and GM3 with thermal-pressure-supported CGM. These three galaxies are ``genetic modifications'' of \po, run with the exact same physics, but starting with slightly perturbed initial conditions that change the galaxies' star formation histories, merger history, and CGM properties (for detailed descriptions of the simulation methods and CGM properties, we refer the interested reader to \citet{Roth:2016, Pontzen:2017, Sanchez:2019}).

GM1 is most like \po\ in that it is actively forming stars at $z = 0.25$ and has a relatively hot CGM. Compared to \po, GM1 has an even stronger accretion flow of cold gas in the inner CGM. In contrast, GM2 and GM3 both quenched around $z \sim 1$ and have significantly less cold gas in their CGM. While these galaxies have different temperature distributions and velocity profiles in their CGM, all three show an offset between the radial velocity distributions of cold and hot gas. While the presence of a kinematic offset between the cold and hot gas is likely a feature of most thermal-pressure-supported CGM around low-redshift L$^*$ galaxies, the quantitative details of that offset, including the presence of accretion streams, will be depend on the details of the galaxy model.

\end{document}